\documentclass{aa}  

\usepackage{graphicx}
\usepackage{txfonts}
\usepackage{placeins}           % useful with \FloatBarrier, to keep 
\usepackage[english]{babel}
\usepackage{natbib}
\usepackage{xspace}
\usepackage{bm}

\usepackage[]{hyperref}
\AtBeginDocument{\nolinenumbers\let\linenumbers\nolinenumbers}

\begin{document}

%\title{This is my last paper and you all are wrong}
\title{The biases caused by interpolated table models and their impact on X-ray spectroscopy}

\author{Matteo Lucchini \inst{1}
      \and
      Benjamin Ricketts \inst{1,2}
      \and 
      Phil Uttley \inst{1} 
      \and 
      Daniela Huppenkothen \inst{1} 
      \and 
      Fergus Baker \inst{3} 
      \and 
      Gregoire Marcel \inst{4}
      \and 
      Guglielmo Mastroserio \inst{5}}

\institute{Anton Pannekoek Institute for Astronomy, University of Amsterdam,
           Science Park 904, 1098XH Amsterdam, the Netherlands
           \and
           SRON, Niel Bohrweg 4, 2033 CA Leiden, the Netherlands 
           \and 
           School of Mathematics, Statistics, and Physics, Newcastle University, Newcastle upon Tyne NE1 7RU, UK
           \and 
           Department of Physics and Astronomy, FI-20014 University of Turku, Finland
           \and 
           Scuola Universitaria Superiore IUSS Pavia, Palazzo del Broletto, piazza della Vittoria 15, I-27100 Pavia, Italy}

\abstract
% context heading (optional)
% {} leave it empty if necessary  
{Models for X-ray spectroscopy involving complex relativistic or plasma physics are commonly pre-computed on a grid, which is then distributed throughout the community and interpolated linearly during a fit.}
% aims heading (mandatory)
{In this paper, we aim to quantify the potential biases and distortions introduced by the computational error and propagated into parameter inference by interpolating linearly these complex, non-linear models.} 
% methods heading (mandatory)
{We analyse the performance of Bayesian sampling algorithms on mock inference problems. Then, we model real observations of an accreting black hole X-ray binary in both the soft and hard states.}
% results heading (mandatory)
{We find that linear interpolation can a) greatly bias the posterior distribution of model parameters and b) distort the likelihood space to such a degree that posterior distributions become artificially multimodal. This multi-modality causes a noticeable loss in computational efficiency in both sampling and fitting techniques. As a result, exploring a given parameter space until convergence requires an extremely large computational cost.}
% conclusions heading (optional), leave it empty if necessary
{These findings call for improving the computational techniques used when fitting modern X-ray data, for example through better sampling algorithms and/or more sophisticated interpolation schemes.}

%\keywords{X-ray astronomy --
%          passive aggressiveness --
%          sheer hatred
%           }

\maketitle

%%%%%%%%%%%%%%%%%%%%%%%%%%%%%%%%%%%%%%%%%%%%%%%%%%%%%%%%%%%%%%
\section{Introduction}

Understanding the X-ray emission of accreting compact objects requires accounting for complex plasma physics \citep[e.g.][]{Poutanen96,nthcomp,Garcia13} as well as relativistic effects \citep[e.g.][]{reflionx,Li05,Dauser13}. As with any complex modeling problem, deriving fundamental physics from fitting data is complicated both by model run-time, and by high-dimensional, correlated, and/or multi-modal parameter spaces. 

It is common to utilize table models to speed up model calculations: a model is pre-computed for a fixed grid of parameter values, and then distributed to the community, who use existing modeling software like \texttt{Xspec} \citep{Arnaud96} to interpolate between the spectra stored in the tables and derive spectra for any choice of parameter values. Examples of commonly used table models are most physical reflection models, such as the \texttt{Relxill/Xillver} \citep{Garcia10,Dauser13,Dauser22} or the \texttt{Reflionx} families \citep{reflionx} of models; relativistic disc models like \texttt{kerrbb} \citep{Li05}, and \texttt{atomdb} plasma emission models like \texttt{apec, aped, vei} or \texttt{cei} \citep{Smith01}. 

Crucially, existing software by default performs linear interpolation. This simplification is unsurprising, as non-linear interpolation in a large (5+) number of dimensions is an extremely complex computational problem. Computing the output value $m$ of a non-linear model through linear interpolation over a parameter $\theta$ inherently introduces an error
\begin{equation}
    e (\theta) =  m(\theta) - \tilde{m}(\theta) = -\frac{1}{2}m^{\prime\prime}(\xi) (\theta-\theta_i)(\theta_{i+1}-\theta), 
    \label{eq:interp_error}
\end{equation}
where $m$ is the true model value, $\tilde{m}$ is the result of the linear interpolation, $\theta_i$ and $\theta_{i+1}$ are parameter values for which the model is pre-computed and stored in the tables, $\theta$ is the point between these two values where the model is being computed, $m^{\prime\prime}(\xi)$ is the second derivative of the model function $m$ evaluated at an appropriate point $\xi$; we derive eq.\ref{eq:interp_error} in Appendix \ref{sec:1d_error}. By definition, if the model is non-linear (e.g., if it is a black body, or power-law) this term is non-zero and therefore the interpolation error is non-zero. 

A standard assumption when one uses a table model in inference is that the interpolation error $e(\theta)$ is negligible - for example, because the uncertainty of the data $\sigma_\mathrm{data}$ dominates when computing confidence intervals. When this assumption is broken, for example in the context of current and future missions with high sensitivity, the parameter credible intervals will be biased by the construction of the tables and their interpolation. The issues that arise when this assumption breaks down have been debated widely in the field of applied mathematics \citep[e.g.][]{Bates80,Kennedy01,Kaipio07}, and are occasionally discussed in the X-ray astronomy  literature \citep[][]{Davis06,Parker19}; they are also mentioned in the \href{https://heasarc.gsfc.nasa.gov/docs/software/xspec/manual/XSappendixLocal.html}{\texttt{Xspec} documentation page} as well as the \href{http://atomdb.org/interpolation/index.php}{\texttt{AtomDB} web page}. However, to date there has not been a demonstration of the impact that these numerical techniques have on the scientific results produced by X-ray spectral fitting, which we aim to provide here. 

The paper is organized as follows. In Sec.\ref{sec:Bayes} we provide an overview of Bayesian sampling; in Sec.\ref{sec:trivial} we derive analytically how interpolation errors affect the chi-squared fit statistic; in Sec.\ref{sec:powerlaw} we build several mock problems to demonstrate how interpolation errors affect posterior distributions; in Sec.\ref{sec:RealObservations} we demonstrate two applications to real data using common fitting methods, using NuSTAR observations of a bright black hole X-ray binary (BHXRB) as an example, and in Sec.\ref{sec:Discussion} we summarise our findings and report our conclusions. All the fits in this paper have been performed using version $0.3.1$ of \texttt{nDspec} \citep{Lucchini26}, which includes support for joint fits, simulating observations, user-defined likelihoods, and interfaces with most Python-based Bayesian sampling packages. A full reproduction package can be found on \href{I 10.5281/zenodo.23564637}{Zenodo}.

\section{An overview of Bayesian sampling}
\label{sec:Bayes}

For many astrophysical problems, models exist that parametrize the physical processes that produce the light we observe. The parameters are usually of fundamental importance to our understanding of the underlying physics in a given system, and thus obtaining accurate and robust estimates of these parameters is imperative. 
One increasingly popular approach is to obtain these estimates through Bayesian statistics. Given a set of $N$ measurements $\bm{d} = \{d_n\}_{n=1}^N$ at positions $\bm{x} = \{x_n\}_{n=1}^N$, and a model $m(x, \theta)$ with parameter vector $\theta$, Bayes' theorem states
\begin{equation}
    p(\bm{\theta} | \bm{d}) = \frac{p(\bm{d}|\bm{\theta})p(\bm{\theta})}{p(\bm{d})} .
    \label{eq:bayes}
\end{equation}
Here $p(\bm{\theta} | \bm{d})$ is the posterior distribution, i.e. the probability distribution of the parameters given the data, $p(d |\bm{\theta})$ is the likelihood function, $p(\bm{\theta})$ is the prior density, and $p(\bm{d})$ is the Bayesian evidence, which acts as a normalisation term for eq.\ref{eq:bayes}. The likelihood describes how observations are generated given the underlying model, while the prior distribution encodes our knowledge of the parameters before we looked at the data. The evidence term is also convenient for model comparison, because it effectively encodes the likelihood of the data given the overall model, integrated over the entire parameter space. However, since calculating $p(\bm{d})$ is computationally intensive and by definition it does not affect the shape of the posterior as a function of $\theta$, it is often neglected for parameter inference. 

Bayesian posteriors are convenient for many astrophysical applications for a number of reasons, because they generally encode the quantity we want to know: the probability distribution over the parameters of interest.  By the rules of conditional probability, they also allow us to integrate over any parameters in the model we do not care about (also called nuisance parameters), a process called marginalization: 

\begin{equation}
 p(\theta_{j} | \bm{d}) = \int_{\Omega_{i\neq j}}  p(\theta_{i\neq j} | \bm{d}) d\theta_{i\neq j}.   
 \label{eq:1d_posterior}
 \end{equation}

The marginalized posterior $p(\theta_j | \bm{d})$ conveniently folds in our uncertainty in the nuisance parameters. Full posteriors are multi-dimensional distributions that are challenging to understand and visualize. Therefore, they are generally summarized in lower dimensions, for example in two-dimensional scatter plots, one-dimensional histograms, or through numerical summaries such as the posterior mean (or, more robustly, median) and percentiles of the one-dimensional marginalized posterior distributions. The latter are known as \textit{credible intervals} and function as a Bayesian equivalent to confidence intervals\footnote{Though their statistical interpretation differs significantly and they should not be confused.}.

However, because for most practical applications Equation \ref{eq:bayes} does not admit an analytical form\footnote{Note that while we generally have analytical forms for the likelihood (e.g. a Gaussian or Poisson distribution) and for the prior, the \textit{product} of these is often not analytically tractable.}, Bayesian inference comes at the cost of having to numerically approximate multi-dimensional integrals. Posterior distributions can in practice be quite complex and non-Gaussian in shape, with sometimes long tails or multiple peaks. In these situations, optimization and peak-finding algorithms may struggle by identifying only one of multiple parameter regions with significant probability mass, and approximation algorithms for one-dimensional parameter uncertainties might miss long tails or multi-peak structures entirely, leading to underestimated uncertainties.

As a result, one often resorts to systematically exploring the parameter space. Because in all but the lowest dimensions, a grid-based approach is entirely infeasible, one must resort to clever quasi-random strategies to explore the multi-dimensional space: ideally, an algorithm should spend more time (and accumulate samples) in high-probability regions, and less time in low-probability regions. A wide range of algorithms exist to tackle that problem, of which the most commonly used in X-ray astronomy is Markov Chain Monte Carlo, and in particular the Affine Invariant Ensemble Sampler (AIES) of \cite{GoodmanWeare2010}, as implemented by \cite{emcee}.

The idea of MCMC algorithms is to create a Markov Chain whose position will, after a sufficient number of iterations, have drawn a reasonable enough number of samples from the posterior distribution in eq.\ref{eq:bayes} that the latter can be approximated to good accuracy. Conveniently, in the case of posterior distributions this implies that marginalising over nuisance parameters simply means constructing a histogram of sampled position for a given parameter $\theta$, ignoring all other parameters. Different MCMC algorithms are differentiated by how they evolve the chain from step $i$ to step $i+1$; how well suited a given step proposal is for a given problem has a large impact on the sampler performance \citep[e.g.][]{zeus,Albert25}.

A key challenge with MCMC algorithms is knowing if and when they sample from the true posterior distribution. While mathematical guarantees exist that an MCMC chain eventually will sample from the posterior, there is no guarantee that it will do so in finite time. For example, a Markov chain that is started far away from any high-probability regions will take a long time to find those high-probability regions, and early samples will not be a good representation of the probability distribution we are interested in. A robust estimate of convergence is the autocorrelation length $\tau$, which encodes how many MCMC steps it takes for a Markov Chain (like a walker in the AIES algorithm) to forget its starting place \citep{GoodmanWeare2010}. Generally, at least one $\tau$ should be discarded as so-called ``burn-in'', where the chain slowly moves towards the stationary posterior distribution. However, until convergence is reached the estimate of $\tau$ is also a function of the total chain length: in order to be sure the estimate of $\tau$ is robust, the chain needs to sample for much longer than a single timescale (a good recommendation is a chain length $\approx$50$\tau$; \citealt{Hogg18}) in order to be somewhat certain the MCMC chain is sampling from the posterior. A second heuristic for sampler health is the acceptance fraction: a fraction that is too low automatically yields extremely long autocorrelation time as the chains are stuck on the same set of parameters for many iterations. A very high acceptance fraction suggests an unconstrained random walk: the sampler is not sampling from the posterior, either. Most modern MCMC samplers, including AIES, are affine invariant, meaning their performance is entirely immune to \textit{linear} correlations between parameters. However, multi-modal (i.e.~multi-peaked) distributions often still present a challenge, especially when these peaks are narrow and well-separated by low-probability troughs. Similarly, they might not handle discrete jumps (or phase transitions) in the likelihood gracefully. 

A second technique commonly used for estimating eq.\ref{eq:1d_posterior} is nested sampling \cite[e.g.][]{Skilling04,Multinest1,Ultranest,Dynesty,Ashton22,Buchner23b}. The idea of nested sampling is to turn the high-dimensional integral computing the normalizing constant $p(\bm{d}) = \int_{\Omega}p(\bm{d} | \theta) p(\theta) d\theta$  into a one-dimensional problem, by shrinking the entire prior volume to the posterior along contours of constant  likelihood. This is achieved through a one-dimensional integral over surfaces of identical likelihood \citep[][]{Multinest1}:
\begin{equation}
    p(\bm{d}) = \int_{0}^{+\infty}X(\lambda)d\lambda = \int_{0}^{+1} \mathcal{L}(X) dX,
\end{equation}
where $\lambda$ is a contour in the parameter space such that the likelihood $\mathcal{L}=\mathcal{}p(\bm{d}|\bm{\theta})=\lambda$, and $X(\lambda)$ is the prior volume contained within that contour, such that the likelihood is $>\lambda$,  or mathematically:
\begin{equation}
    X(\lambda) = \int_{\bm{\theta}:p(\bm{d}|\bm{\theta})>\lambda} p(\bm{\theta})d\bm{\theta}.
\end{equation}

In the initial sampling step, samples are drawn randomly from the prior, constituting a population of ``live points''. This population of live points is updated at each successive step by discarding the live point with the lowest likelihood $\lambda$ and resampling a new point with likelihood $p(\bm{d}|\bm{\theta})>\lambda$. This procedure continues until the volume covered by the live points is a small fraction of the starting prior volume. When this ``small enough'' fraction of the initial volume is reached, the sampler is said to have converged to the target posterior distribution. Nested Sampling was generally developed to approximate the evidence integral. However, through importance resampling, the (live and discarded) points sampled throughout this procedure can be transformed to approximate the posterior distribution with little additional computational effort. Various variants of Nested Sampling differ in how contours of equal likelihood are drawn, and in how new live points are generated, among others.

\section{Likelihood distortions}
\label{sec:trivial}
%really this should be titled "first year calculus" but eh

Before testing different samplers, it is useful to examine the analytical form of the likelihood when an interpolation error is present. Given a  model $m(\theta)$, the Gaussian log-likelihood (commonly denoted $\chi^2$) is:
\begin{equation}
    \ln\mathcal{L}(\theta) = -\frac{1}{2}\sum_{k}{\left(\frac{d_k - m_k(\theta)}{\sigma_k}\right)}^2,
    \label{eq:chi-squared}
\end{equation}
where for each bin $k$, $d_k$ is the data point, $\sigma_k$ its error, and $m_k(\theta)$ is the model value in that bin. Let us assume that this particular problem is uni-modal, such that there is a single value $\theta_{\rm true}$ which optimises eq.\ref{eq:chi-squared}. When using a tabulated version $\tilde{m}(\theta)$ of the same model over some pre-computed grid in a single parameter $\theta$, the likelihood becomes:
\begin{equation}
\ln\tilde{\mathcal{L}}(\theta) = -\frac{1}{2}\sum_{k}{\left(\frac{d_k - m_k(\theta) + e_k(\theta)}{\sigma_k}\right)}^2,
\end{equation}
where $e_k(\theta)$ is the interpolation error introduced in eq.\ref{eq:interp_error}. Expanding the square terms:
\begin{equation}
\ln\tilde{\mathcal{L}}(\theta) = \ln{\mathcal{L}}(\theta) - \frac{1}{2} \sum_{k} \frac{e_k^{2}(\theta)}{\sigma_k^{2}} - \sum_{k}{\frac{e_k(\theta)[d_k-m_k(\theta)]}{\sigma_k^{2}}} ; 
\label{eq:table_likelihood}
\end{equation}
in this form, we see that the table model introduces two additional terms, which notably only vanish at the grid points. The first term shifts the likelihood by some amount which depends only on the distance from the grid points and the grid construction; therefore, it penalises (or favours, depending on the sign of the model derivative) the fit as $\theta$ moves away from a grid point. The second term can never be zero, even if the model is perfectly specified (meaning it fits the data perfectly); due to statistical noise and inherent uncertainty in the underlying parameters, the sum $d_k-m_k(\theta)$ over all bins will never be exactly zero. It can bias the fit in cases in which the residuals $d_k-m_k(\theta)$ have a structure that is similar to the interpolation error $e_k(\theta)$, such that the sum $\sum_k e_k(\theta)[d_k-m_k(\theta)]$ is large.

Let us now examine the gradient of the likelihood $\ln\tilde{\mathcal{L}}(\theta)$ with respect to the interpolated parameter $\theta$:
\begin{equation}
    \frac{d \ln\tilde{\mathcal{L}}}{d \theta} = \frac {d \ln{\mathcal{L}}(\theta)}{d \theta} + \frac{d}{d\theta}\left( -\frac{1}{2} \sum_{k} \frac{e_k^{2}(\theta)}{\sigma_k^{2}} - \sum_{k}{\frac{e_k(\theta)[d_k-m_k(\theta)]}{\sigma_k^{2}}}  \right);
\end{equation}
from eq.\ref{eq:interp_error} and (to leading order in the grid spacing) treating $m^{\prime\prime}(\xi)$ (with $\xi$ defined as in \ref{sec:1d_error}) as constant  across the grid cell:
\begin{equation}
    \frac{d e(\theta)}{d\theta} \approx \frac{1}{2}m^{\prime\prime}(\xi)[2\theta-\theta_i-\theta_{i+1}] \equiv e^{\prime}(\theta), 
    \label{eq:error_gradient}
\end{equation}
and therefore, by applying the chain rule, re-arranging the terms, and dropping the explicit dependence on $\theta$ for brevity, we have:
\begin{equation}
 \frac{d \ln\tilde{\mathcal{L}}}{d \theta} = \frac {d \ln{\mathcal{L}}}{d \theta} - \sum_{k} \frac{e_k e_k^{\prime} - e_k m^{\prime}_k+ e_k^{\prime}(d_k-m_k) }{\sigma_k^{2}}.  
 \label{eq:gradient}
\end{equation}
The three additional terms have different behaviour around grid points. The first term, $e_k e_k^{\prime}$, vanishes at the grid points as well as at the middle point $\theta=(\theta_i+\theta_{i+1})/2$. As a consequence of having three roots, this term introduces a bias in the gradient which changes sign within a single grid interval. On the other hand, the second term $e_k m^{\prime}_k$ has a constant sign between two grid points (as long as $m^{\prime}_k$ does); the bias it introduces is only zero at a grid point, or $\theta=\theta_i$ for any $i$. Finally the third term, $e_k^{\prime}(d_k-m_k)$, by definition introduces a discontinuity when moving $\theta$ between grid points (eq.\ref{eq:error_gradient}). It introduces a discrete jump in the gradient, which can cause gradient-based methods to fail. Additionally, if there are values of $\theta$ such that:
\begin{equation}
\frac {d \ln{\mathcal{L}}}{d \theta} = \sum_{k} \frac{e_k e_k^{\prime} - e_k m^{\prime}_k+ e_k^{\prime}(d_k-m_k) }{\sigma_k^{2}}   
\end{equation}
the linear interpolation will introduce additional, artificial modes in the posterior. From these equations, we see that linear interpolation can both bias the posterior, and introduce artificial solutions that need not be consistent with or close to $\theta_{\rm true}$.

\begin{figure*}
    \centering
    \includegraphics[width=\textwidth]{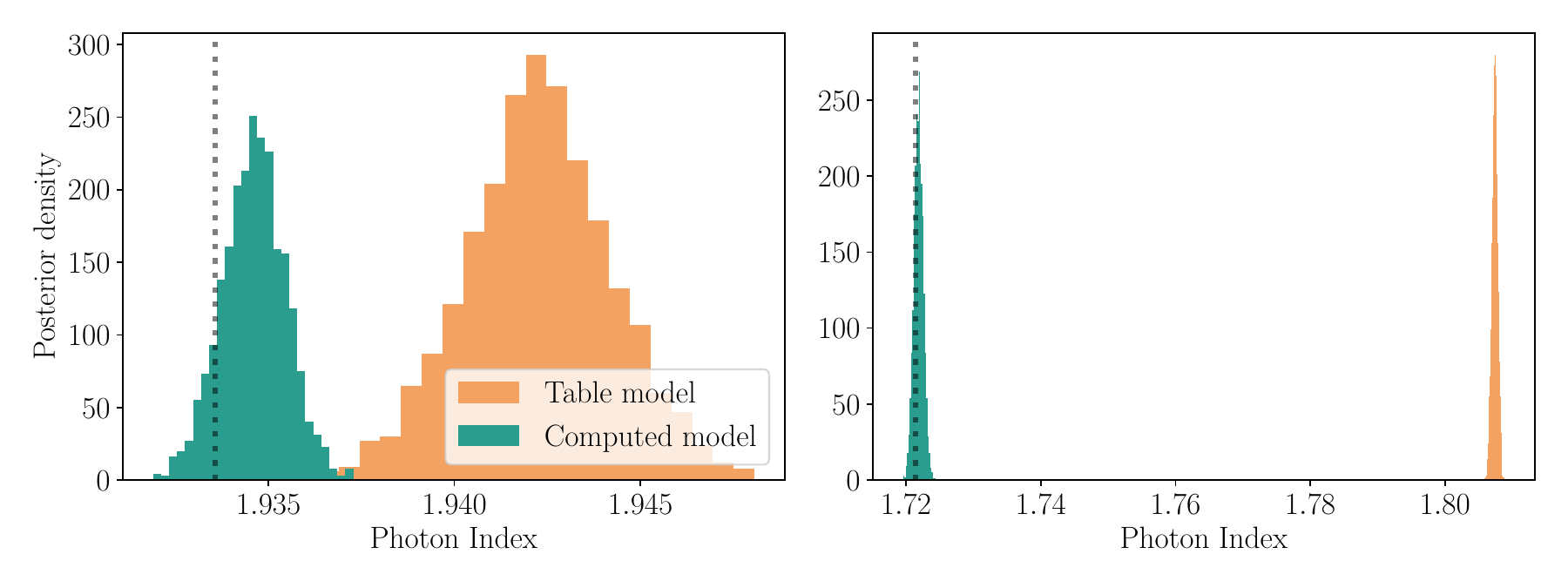}
    \caption{Posterior distribution when the true value is near a grid point and the interpolation error is minimal (left plot), and in between two when the interpolation error is maximal (right plot). The true value is shown by the dashed line.}
    \label{fig:pl_posterior}
\end{figure*}

The corresponding $n-$dimensional formulae for eq.\ref{eq:interp_error}, eq.\ref{eq:table_likelihood} and \ref{eq:gradient}, which we derive in Appendix \ref{sec:nd_derivation}, are as follows.
\begin{equation}
    e_k(\bm{\theta}) = -\frac{1}{2}\sum_j \frac{\partial^{2}m_k(\bm{\theta})}{\partial \theta_j^{2}}(\theta_j-\theta_{j,i})(\theta_{j,i+1}-\theta_j),
    \label{eq:nd_error}
\end{equation}
where $e(\bm{\theta})$ is the total error introduced by interpolating over each parameter $\theta_j$, and $\theta_{j,i}$ and $\theta_{j,i+1}$ are grid points identically to eq.\ref{eq:interp_error}.
\begin{equation}
\ln\tilde{\mathcal{L}}(\bm{\theta}) = \ln{\mathcal{L}}(\bm{\theta}) -\sum_k\sum_j \frac{e_{k}^{(j)}(d_k-m_k)}{\sigma_k^{2}} -\frac{1}{2}\sum_k\sum_j\sum_l \frac{e_{k}^{(j)}e_{k}^{(l)}}{\sigma_k^{2}}, 
\label{eq:nd_likelihood}
\end{equation}
where $e_{k}^{(j)}$ is the $j$-th component of the total error $e_k(\bm{\theta})$ in a given bin $k$, contributed only by interpolating over parameter $\theta_j$. 
\begin{equation}
\frac{\partial \ln\tilde{\mathcal{L}}}{\partial \theta_j} = \frac {\partial \ln{\mathcal{L}}}{\partial \theta_j} - \sum_{k} \frac{e_k \epsilon_k^{(j)} - e_k \partial m_k/\partial \theta_j+ \epsilon_k^{(j)}(d_k-m_k) }{\sigma_k^{2}},  
\label{eq:nd_gradient}
\end{equation}
where the derivative of the error
\begin{equation}
    \epsilon_k^{(j)}  \equiv  \frac{\partial e_k(\bm{\theta)}}{\partial \theta_j} = \frac{1}{2}\frac{\partial^{2} m_k}{\partial \theta_j^{2}}(2\theta_j-\theta_{j,i}-\theta_{j,i+1})
\label{eq:nd_deriv}
\end{equation}
is analogous to eq.\ref{eq:error_gradient}, computed with respect to parameter $\theta_j$.

Compared with the one-dimensional case, two changes appear. First, in eq.\ref{eq:nd_likelihood} cross terms between interpolation errors appear through the quadratic term $e_{k}^{(j)}e_{k}^{(l)}$. Second, the terms $\propto e_k$ in eq.\ref{eq:nd_gradient} mean that the gradient of the likelihood with respect to parameter $\theta_j$ depends on the interpolation error on all parameters, not just $\theta_j$. Intuitively, this shows that the interpolation errors become progressively more important, and distort the parameter space more, as one interpolates over more parameters. As a result, a given model likelihood (and therefore, a given posterior distribution) can be impacted by linear interpolation errors in one of three ways.

First, let us consider a case in which the posterior is sufficiently narrow not to cross any grid points. In this regime, the additional interpolation terms simply contribute a near-constant bias which will move the credible interval away from the correct range. The larger the interpolation error (because of either a coarse grid, a true parameter value far from the grid points, or both), the more important this effect becomes. 

Second, let us assume that the term $e_k[d_k-m_k]$ is dominant. This can happen either because the model is misspecified, leading to large structured residuals $d_k-m_k$, or because one is interpolating over many parameters and/or using a coarse grid, leading to large $e_k$. Let us also assume that the posterior peak is very near, or exactly on, the grid point $\theta_i$. In this case, the additional likelihood terms are zero near the true peak, and moving away from it depends on the sign of $e_k\propto \partial^{2}m(\theta)/\partial\theta^{2}_j$ (eq.\ref{eq:error_gradient} and \ref{eq:nd_deriv}). If the sign is positive, then the likelihood improves as one moves away from  $\theta_i$, and the result is a bi-modal distribution centred on the grid point. If the sign is negative, then the posterior is narrowed artificially, as the error from interpolation lowers the likelihood.

Third, let us assume that the model fits the data well, such that the term $d_k-m_k$ is small, but the posterior is broad enough to cross multiple grid points; this can happen, for example, because the parameter $\theta_i$ is degenerate with some other model parameter. In this regime, the quadratic terms $e_{k}^{(j)}e_{k}^{(l)}$ become dominant away from grid points, and vanish to zero at the grid points. The result is a series of peaks and valleys in the posterior, each corresponding to a grid point. In extreme cases, these can become separated into multiple well-defined modes.

\section{Sampler behaviour in example cases}
\label{sec:powerlaw}

In this section, we illustrate each of the three regimes highlighted in sec.\ref{sec:trivial} with three straightforward examples, using  absorbed power-laws or colour-corrected discs as continuum models. We modify either the data or model as necessary to highlight the different terms in eq.\ref{eq:table_likelihood}, and compare using a model computed on the fly, versus a table model. Additionally, we compare the loss in performance due to sampling higher dimensions (the ``curse of dimensionality''), with that caused by sampling multi-modal distributions that linear interpolation can cause. 

\subsection{Regime one: biased posterior recovery}
\label{sec:regime_one}

In this first example, we utilise an absorbed power-law as a toy model. Choosing such a simple one-dimensional model is particularly useful, because the error terms in the likelihood can be estimated analytically. Most table models used in the community are both high-dimensional in parameter space and computationally expensive, such that the number of models that can be produced is limited. As a consequence, the grids in most table models are very coarse in each dimension, and estimating the interpolation error is unfeasible. Consequently, we tabulate the model for only eight linearly-spaced photon indexes between $\Gamma=1.1$ and $\Gamma=4$. 

We simulate two spectra: one with photon index near one of the grid points ($\Gamma=1.9335$, while the nearest grid point has $\Gamma=1.929$), and one in-between two grid points ($\Gamma=1.7214$, versus $\Gamma=1.514$ for the other nearest grid value). Additionally we set the normalisation to $N=5$, the column density to $N_H=0.1\times10^{22}$ cm$^{-2}$, and simulate 20 ks observations with NuSTAR using the \texttt{simulate\_spectrum} function in \texttt{nDspec}. This function convolves a model with an instrument response, re-scales it by exposure time, and generates a Poisson realisation of the expected spectrum, identically to the \texttt{fakeit} command of \texttt{Xspec}. 

\begin{figure}
    \centering
    \includegraphics[width=\columnwidth]{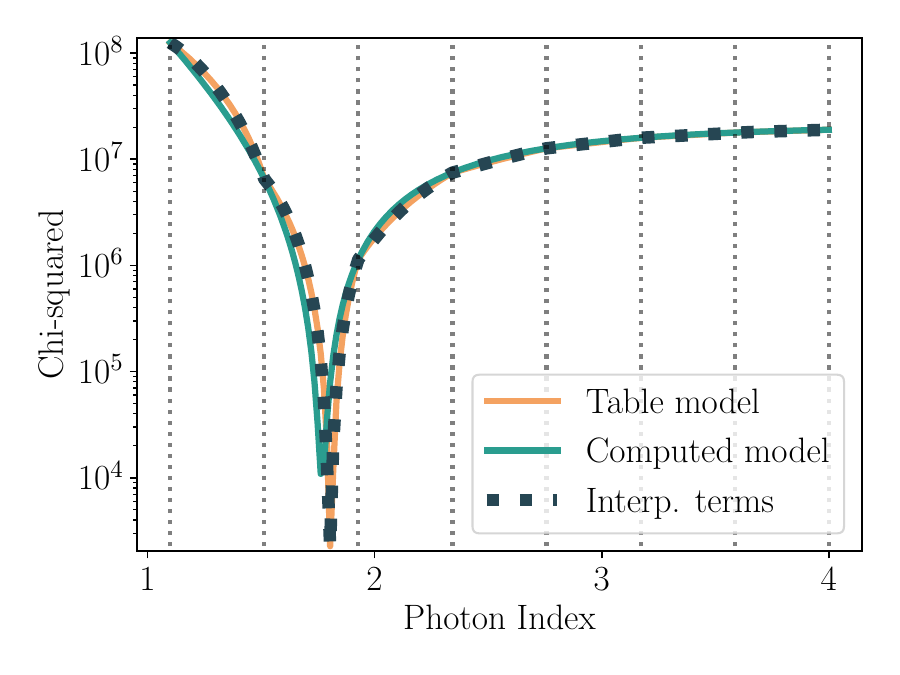}
    \includegraphics[width=\columnwidth]{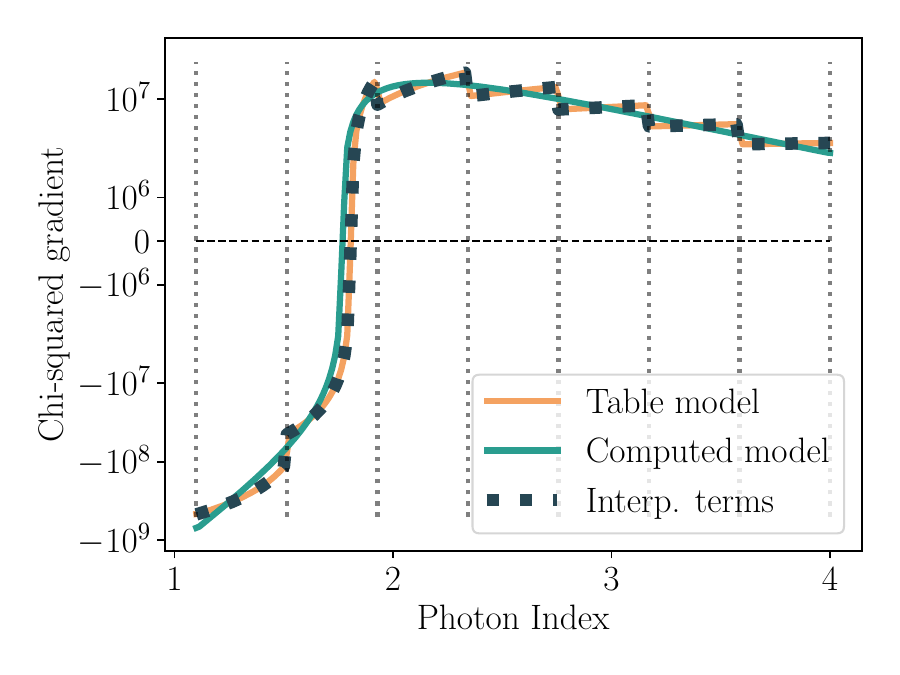}
    \includegraphics[width=\columnwidth]{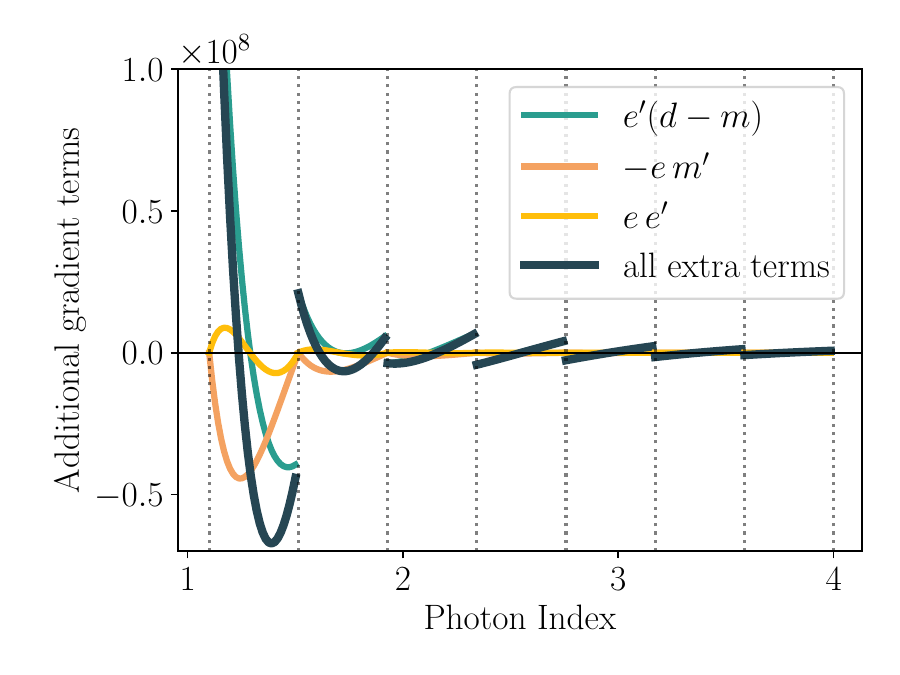}
    \caption{Top panel: fit statistic as a function of the photon index. Middle panel: the gradient of the fit statistic. Bottom panel: contribution of the terms in eq.\ref{eq:gradient} to the likelihood gradient. The grid of photon indexes is shown by the dotted lines.}
    \label{fig:pl_fitstats}
\end{figure}

For all bins, the number of photons per bin is $\approx$tens to thousands, dropping at high energies, and therefore using a Gaussian likelihood is reasonable approximation. We fit the simulated data between $3$ and $60$ keV by optimizing the $\chi^{2}$ fit statistic, and then use the optimisation solution to initialize an AIES sampler. We use flat priors with boundaries at $[1.1,4]$, $[0.1,10]$ and  $[0.05,0.4]$ for the photon index, normalisation, and column density respectively. We run the sampler for $2500$ steps; at the end of the run, we find an autocorrelation length $\tau\approx 35$ steps for all three model parameters, and therefore consider the chain converged to the true posterior. We produce posteriors by taking $2\max(\tau)$, twice the largest autocorrelation length, as the burn in time, and by thinning the chain by a factor equal to $\min(\tau)$. 

Fig.\ref{fig:pl_posterior} shows a comparison of the posteriors, in the cases when the true parameter value is close to (left panel) or far from (right panel) a pre-computed grid point. As expected from sec.\ref{sec:trivial}, we find a bias in the parameter recovered using the table model. In the best case scenario (left plot), when the true value is close to the grid and therefore the error introduced by the interpolation is minimal (eq. \ref{eq:interp_error}), we still notice a bias on the order of a few $\sigma$. If the true value lies far from the pre-computed grid (right plot), then the table model fails catastrophically in recovering the correct parameter value, with a bias on the order of $\geq10\sigma$. 

The reason for this failure in recovering the true value becomes apparent in Fig.\ref{fig:pl_fitstats}, where we compare the value of the $\chi^{2}$ statistic, its gradient, and the additional terms in eq.\ref{eq:gradient}, as a function of $\Gamma$. For the table model, the maximum likelihood is at a value of $\Gamma$ that is systematically offset from the true value (top panel). Additionally, the fit statistic computed for the table model is not a smoothly curved function: it is closest to the on-the-fly model near the grid points, but it becomes worse for values of $\Gamma$ that are farthest from the pre-computed grid (middle panel). These discontinuities and bumps are caused entirely by the additional terms introduced by linear interpolation (bottom panel). Indeed, when we modify the value of the fit statistic of the computed model by the terms in eq.\ref{eq:table_likelihood} and eq.\ref{eq:gradient}, shown by the dotted dark grey line, we recover the fit statistic (and its gradient) of the table model exactly. This example, therefore, reproduces the first of the three regimes discussed in the previous section.

Note that, in this case and this case alone, one can completely rectify the problem with logarithmic interpolation alone (see Appendix \ref{sec:log_interp}), but every table model should be inspected for the above discrepancies either analytically or numerically.

\subsection{Regime two: crossing a single grid point}
\label{sec:regime_two}

In this second example, we use the same model and methodology as in the first example, but we purposefully introduce model misspecification. The goal of this test is to verify how the last term in eq.\ref{eq:table_likelihood} can impact the posterior distribution of a tabulated model parameter (in this case, the photon index). 

We do so by introducing a small amount of curvature in the simulated spectrum, which now follows a log-parabola: $d(E) \propto E\exp[-\Gamma+\beta\ln(E)]$; we take $\beta=0.03$. The additional energy dependence ensures that the term $\sum_k e_k(d_k-m_k)$ is large. We keep the same column density and power-law normalisation as the previous example, but change the photon index such that the best-fit value found with both computed and table models lie close to a grid point. In particular, we found that taking $\Gamma=2.038$ produced fits near the grid point $\Gamma_i=1.9285$. We fit the data identically to Sec.\ref{sec:regime_one}. 

\begin{figure}
    \centering
    \includegraphics[width=\columnwidth]{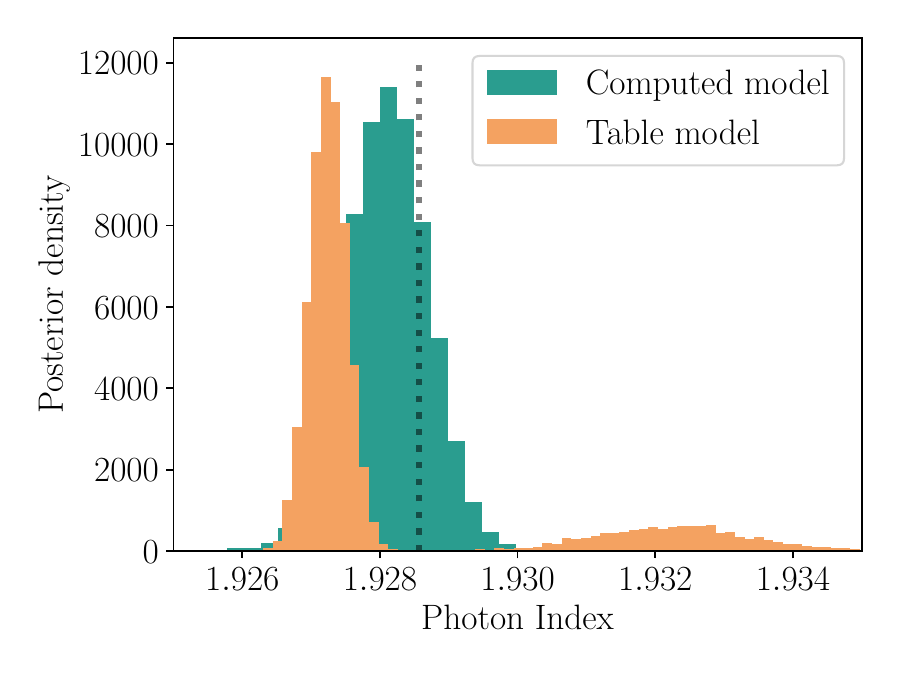}
    \caption{Posterior distributions with large model misspecification. In the case of the table model, the posterior becomes artificially bi-modal. The table model grid point is shown by the dotted line.}
    \label{fig:bimodal}
\end{figure}

We then compare again the posterior distribution of the photon index produced by the table and computed models. These are shown in Fig.\ref{fig:bimodal}. We see immediately that the posterior produced with the table model is bi-modal, with a valley at exactly the grid point, just as we calculated in Sec.\ref{sec:trivial}. No such feature is seen by the computed model, which simply produces a uni-modal posterior. We also note that both models are biased from the true solution $\Gamma\approx2.01$, as a result of model misspecification, and that the bias discussed in Sec.\ref{sec:regime_one} is still present. Finally, we find that the performance of the sampler degrades somewhat for the table model: the acceptance fraction decreases from $\approx 60\%$ to $\approx 45\%$, and the autocorrelation length increases from $\approx 35$ to $\approx 50$.

\subsection{Regime three: crossing several grid points}
\label{sec:regime_three}

In this third example, we illustrate a case in which the posterior of a parameter crosses over multiple grid points. The goal of this test is to verify the impact of the second term of eq.\ref{eq:table_likelihood} on the posterior distribution, using a case in which significant model degeneracy is present.

The model we use is a colour-corrected \texttt{diskbb}, in which the model is tabulated over colour-independent energies, $E_{\rm tab} = E/f_{\rm col}$, and for a given effective temperature $T_{\rm eff}$ the flux is $D(E,T_{\rm eff}) \propto f_{\rm col}^{-2}D(E_{\rm tab},T_{\rm eff})$. This implementation is identical to that in \texttt{Xspec} relativistic disc models like \texttt{kerrbb} or \texttt{bhspec}. The grid of disc temperatures has linearly spaced points $0.1$ keV wide, between $0.5$ and $4$ keV. As in the previous examples, we simulate a $20$ ks NuSTAR observation; the true model parameters are $nH=0.1$, $T_{\rm eff}=1.5$ keV, $f_{\rm col}=1.7$, and $\log_{10}(\rm{Norm_{disc}})=1.7$. 

\begin{figure}
    \centering
    \includegraphics[width=\columnwidth]{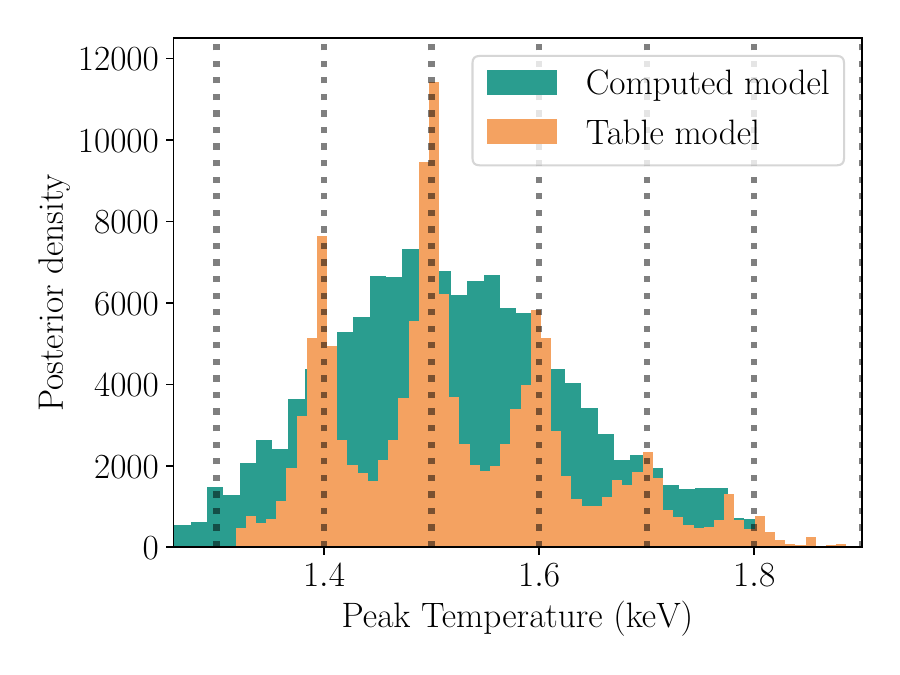}
    \caption{Posterior distributions crossing multiple grid points, showing multi-modality due to the linear interpolation error. The table model grid point is shown by the dotted line.}
    \label{fig:multimodal}
\end{figure}

As in the previous example, we optimise the fit statistic and then reconstruct the posteriors with an AIES run, this time using $5000$ steps. We use uniform priors for $nH$ and $T_{\rm eff}$, log-uniform priors for $\rm{Norm_{disc}}$, and Gaussian prior for $f_{\rm col}$ with mean $\mu=1.7$ and width $\sigma=0.15$. Similarly to the previous case, we find that sampling the table model posteriors is more inefficient, with acceptance fractions dropping from $\approx 30\%$ to $\approx20\%$. The posteriors of the effective temperature reconstructed with both models are shown in Fig.\ref{fig:multimodal}. 

We see that the computed model produces a wide and relatively flat distribution over a large range of temperatures. This is not surprising, as the temperature is highly degenerate with both the disc normalisation $\log_{10}(\rm{Norm_{disc}})$ and the colour-correction factor $f_{\rm col}$. On the other hand, the posterior from the table model remains wide, but it also shows discrete peaks around the model grid points. Once again, this behaviour is exactly that predicted by eq.\ref{eq:table_likelihood}. 

\subsection{Higher dimensions and the curse of dimensionality}
\label{sec:curse}

We now compare the impact of the multi-modality we identified, with an increasing number of model parameters, on sampling performance. It is well established that most sampling algorithms degrade in performance in the case of highly dimensional parameter spaces; this is the problem known as the curse of dimensionality \citep[e.g.][]{Roberts01}. We note that the curse of dimensionality does not affect all problems equally, and depending on the complexity of the parameter space, it may be more or less damaging for sampler performance \citep{Albert25}. 

In order to illustrate how the curse of dimensionality combines with distortions in the likelihood space, we set up two ad hoc problems, and then compare the performance of the AIES sampler as a function of number of dimensions. In all cases, we use 32 walkers and 5000 steps. In the first problem, we sample uncorrelated\footnote{Since AIES is immune to linear correlations between parameters, introducing a covariance term has no impact on the sampler performance.} one-dimensional Gaussian distributions along each dimension, with mean and standard deviation generated randomly between 0 and 1. In the second problem, we complicate the likelihood space by sampling a multi-modal distribution, using a mixture model of three Gaussians. The width of the Gaussians, as well as the mean of the first Gaussian, are generated identically to the first problem. The means of the second and third components are shifted up/down by one. In this way, we ensure that the three Gaussians overlap enough that it is possible for the sampler to identify them.

\begin{figure}
    \centering
    \includegraphics[width=\columnwidth]{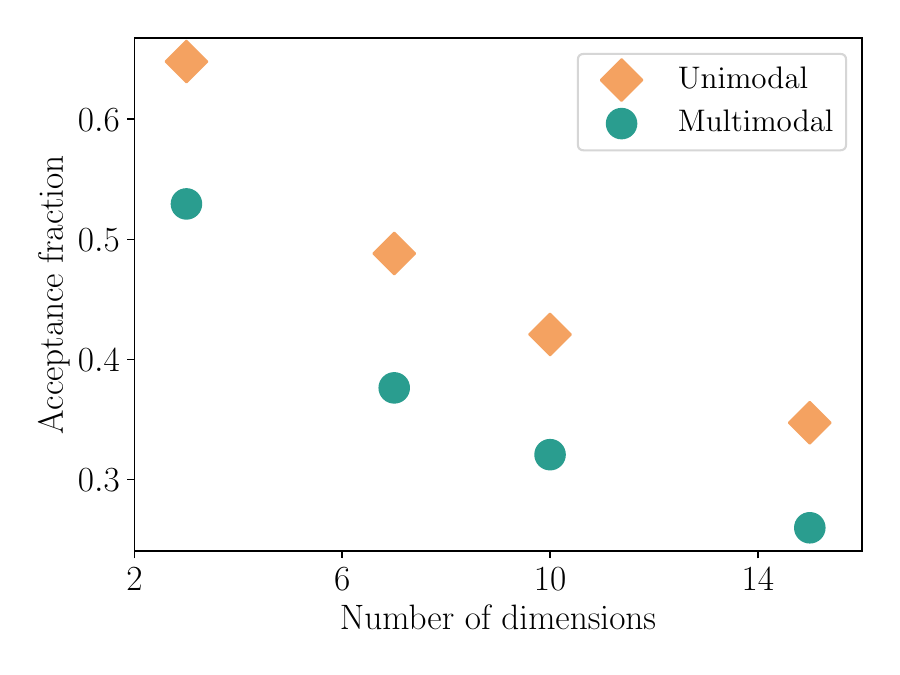}
    \caption{Acceptance rate of different chains, as a function of problem posterior shape and dimensionality. A unimodal, moderately dimensional problem $n=15$ is as challenging for the sampler as a multimodal, low dimensional $n=7$ one.}
    \label{fig:fitaccept}
\end{figure}

For each chain, we compute the acceptance fraction at the end of each run as a proxy for the performance of the sampler; we plot them for both problems in Fig.\ref{fig:fitaccept}. Two things are immediately evident: first, the effect of the curse of dimensionality is obvious and follows the expected shape \citep{Roberts01}, as the acceptance fraction decreases at higher dimensions. Second, and more importantly, the type of problem has a very large impact as well, and sampling a multimodal distribution is far more difficult. In fact, the sampler achieves (roughly) identical acceptance fraction $\approx37\%$ in the uni-modal case in 15 dimensions, as in the multi-modal case in 7 dimensions. 

These two problems combine to greatly increase the difficulty in retrieving meaningful parameter estimates with current X-ray spectroscopic models, software and techniques. 

\section{Sampler behaviour in real observations}
\label{sec:RealObservations}

In this section we analyse a set of standard NuSTAR spectra of the black hole X-ray binary MAXI J1820+070 in both the corona-dominated hard state, and the disc-dominated soft/soft intermediate state. In particular, we demonstrate that the table models typically employed to fit these observations have parameter spaces so badly behaved that retrieving meaningful posterior estimates is very challenging, even when employing methods that ignore the likelihood gradient. 

\subsection{Data reduction}

We reduce the data identically to \cite{Buisson19}. In particular, we select a single ObsID for the hard state (90401309004) and one for the soft-intermediate state (SIMS, obsid 90401309025), in order to test different models that are appropriate for different accretion regimes. We use \texttt{Nupipeline} version 0.4.9 and adopt identical screening criteria to \cite{Buisson19}: "saamode" was set to "strict" and "tentacle" to "yes". Due to the high count rate in these observations, we prevent source photons from being flagged as "test" with the status expression "STATUS==b0000xxx00xxxx000". For each FPM we take regions of 60 arcseconds centred on the source to extract the spectrum, and extracted away from the source for the background. We group the spectra to contain at least 50 counts per channel and ignore channels at high energies that do not have sufficient count to meet this criterion. This results in our spectra extending to 77.5 keV in the hard state, and 73 keV in the SIMS. 

\begin{table}[]
    \centering
    \footnotesize{
    \begin{tabular}{c|c}
        Parameter & Prior  \\ \hline
        Photon index $\Gamma$ & Uniform $[1.1,2.6]$ \\
        $E_{\rm cut}$ (keV) & Uniform $[20,200]$ \\
        Powerlaw normalisation & Uniform $[0.01,10]$ \\ \hline
        Spin $a$ & Uniform$^{*}$ $[-1,1]$ \\   
        Inclination (deg) & Normal$^{**}$ mean=$65^{\circ}$, $\sigma=4^{\circ}$ \\    
        Accretion rate $\dot{M}$ ($\dot{M}_{\rm Edd}$) & Uniform $[0.01,10]$ \\
        Distance $D_{\rm bh}$ (kpc) &  Normal$^{**}$ mean=$3$, $\sigma=1.5$ \\ \hline
        Black body $kT$ (keV) & Uniform $[0.5,3]$ \\
        Black body normalisation & Log-uniform $[-2,5]$ \\ \hline
        FPMB cross-calibration & Uniform $[0.75,1.25]$ \\ \hline
    \end{tabular}
    }
    \newline
    $^{*}$ for the run with fixed spin we take $a=0.75$
    \newline
    $^{**}$ From \cite{Atri20}
    \caption{Model parameters and priors used to fit the SIMS data}
    \label{tab:kerrbb}
\end{table}
\subsection{BHXRB spin measurements in the SIMS}
\label{sec:softstate}

We fit the data using the same model as \cite{Fabian20}: a combination of a relativistic disc using \texttt{kerrbb}, a power-law excess using a cutoff powerlaw model, and an additional black body component which \cite{Fabian20} associate with the emission from the plunging region. Here we do not claim any particular physical interpretation from the data, and aim exclusively at building a deeper understanding of the parameter space of such a model.

We begin by optimising the $\chi^{2}$ fit statistic, with the goal of finding a reasonable starting point for the AIES sampler. The fit statistic optimizer finds a reduced $\chi^{2}/{\rm d.o.f.}\approx 1.1$; for the purpose of this work, this constitutes a sufficiently good description of the data to continue our analysis. After finding a reasonable initial guess, we initialize 10 walkers per parameter (for a total of 100 when the spin is fixed, and 110 when it is free) within 1\% of the best fit found with the optimizer and run the chain for 150000 steps. We use uniform priors for all parameters, except for a Gaussian prior for the inclination taken from optical data \citep{Atri20}. The full list of parameters and their priors is in Tab.\ref{tab:kerrbb}. Identically to Sec. \ref{sec:powerlaw}, after the chain is completed we compute the autocorrelation lengths $\tau$ for each parameter; we take the burn-in to be $2\max(\tau)$ and thin the chain by a factor $\min(\tau)$ before computing the posteriors and trace plots. 

\begin{figure}
    \centering
    \includegraphics[width=\columnwidth]{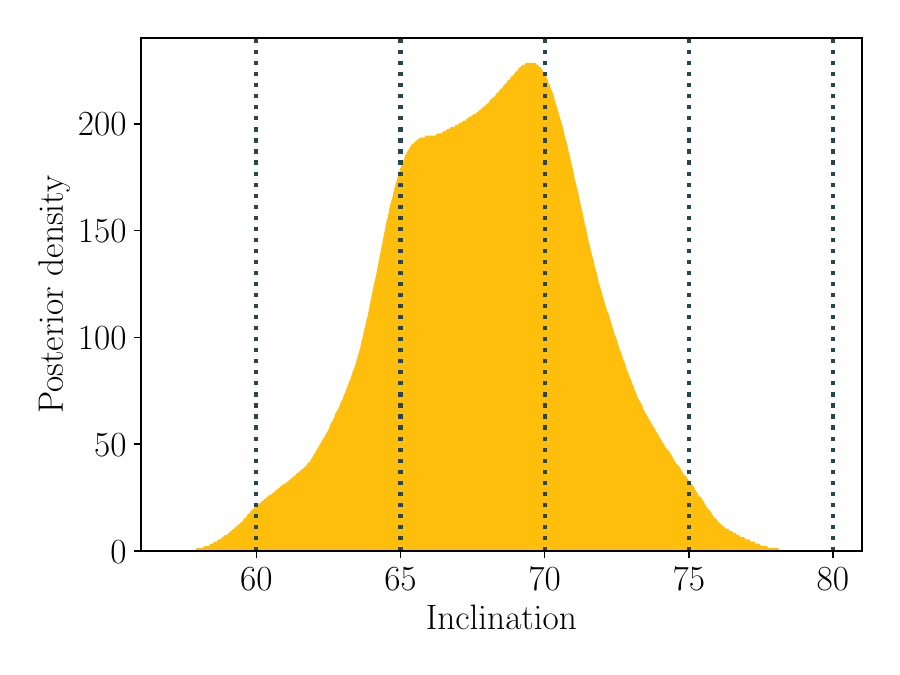}
    \includegraphics[width=\columnwidth]{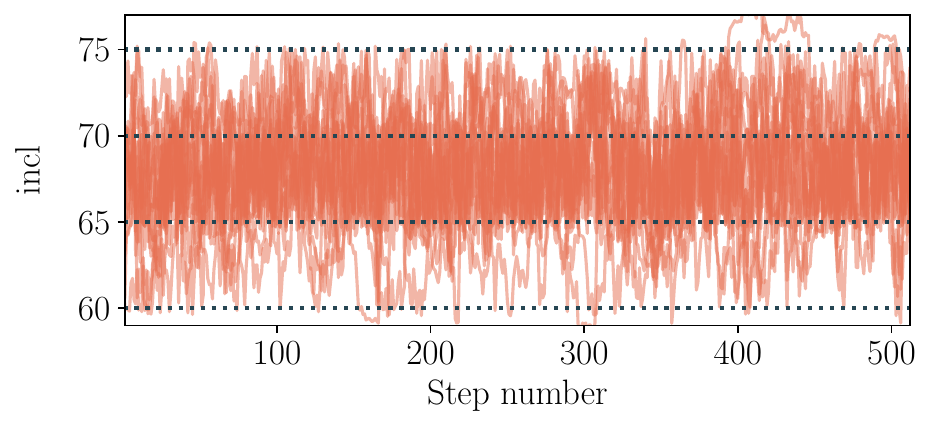}
    \caption{Top panel: one-dimensional posterior for the source inclination. Bottom panel: trace plot for the same parameter. Some of the walkers appear to cluster or be bounded by grid points, which are shown by the dashed lines.} 
    \label{fig:softstate_inclination}
\end{figure}

We begin by fixing the spin to 0.75; in this case, the sampler seems to behave reasonably well. The average acceptance fraction among all walkers is $\approx17\%$, which is somewhat low for a 10-dimensional problem (see Fig.\ref{fig:fitaccept}) but not alarming. We find $\tau\approx500-2000$ depending on the model parameter, which is sufficiently short that we can claim this chain converged to the posterior distribution. Upon more careful analysis, however, even this relatively benign case shows some troubling behaviour, which explains the somewhat low efficiency of the sampler despite the relative simplicity of the problem. This is best illustrated by examining the posterior distribution and trace plot of the inclination, shown in  Fig.\ref{fig:softstate_inclination}. The posterior shows a unimodal distribution, but it also shows a broad plateau between two grid points over which the model was pre-computed. The trace plot shows further evidence that the interpolation and grid are affecting the sampling: we see that in the region between 65 and 70 degrees the walkers have a higher density; in the intervals between 60-65 and 70-75 the density drops abruptly, and they struggle to sample outside of the grid points at 60 and 75 degrees. These are hints that this posterior distribution is in the regime discussed in Sec.\ref{sec:regime_three}. 

We then re-run the sampler on the same problem, but including the spin as a free parameter. This single change has a disastrous impact on the sampler performance. For the same chain length we find an average acceptance fraction of about 5\%, while the estimated $\tau$ increases to between 4000 and 16000 steps depending on the parameter. Furthermore, this estimate should be regarded as a lower limit, because our chain has not run enough to ensure convergence, meaning that the true value $\tau$ is severely under-estimated \citep{Hogg18}. This degradation in performance for a single additional dimension is too severe to originate purely from the curse of dimensionality (sec.\ref{sec:curse}).

\begin{figure}
    \centering
    \includegraphics[width=\columnwidth]{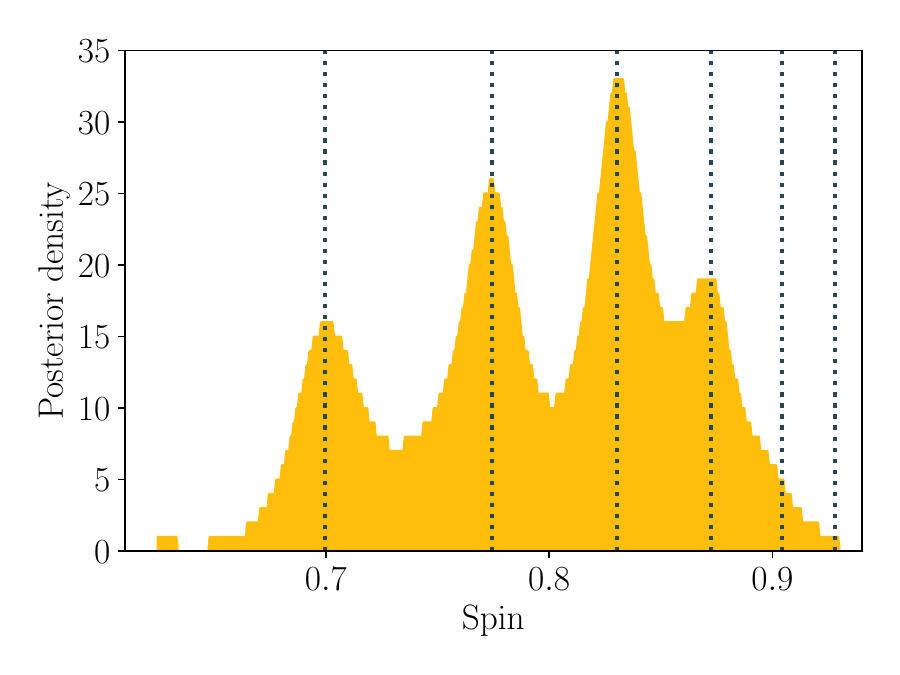}
    \includegraphics[width=\columnwidth]{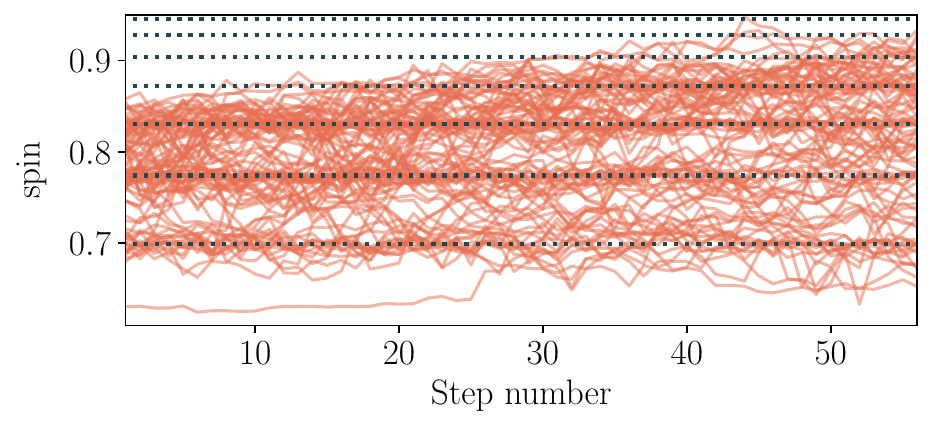}
    \caption{Top panel: one-dimensional posterior for the spin. Bottom panel: trace plot for the same parameter. The non-physical impact of the grid points (in dashed lines) on the walker paths, and therefore on the posterior, is obvious.}
    \label{fig:spin_posterior}
\end{figure}

Inspecting the posteriors and trace plots once again (Fig.\ref{fig:spin_posterior}) immediately explains the cause for the sampler behaviour: we see that the posterior for the spin is highly multimodal, with the peaks corresponding to each of the spin grid points (top plot in Fig.\ref{fig:spin_posterior}). This is exactly the regime identified in Sec.\ref{sec:regime_three}. The multimodal shape of the posterior introduced by the fixed-grid table model makes it challenging for MCMC walkers to mix efficiently, and they require many iterations to diffuse into all of the relevant modes (if they manage to at all). 

\begin{figure}
    \centering
    \includegraphics[width=\columnwidth]{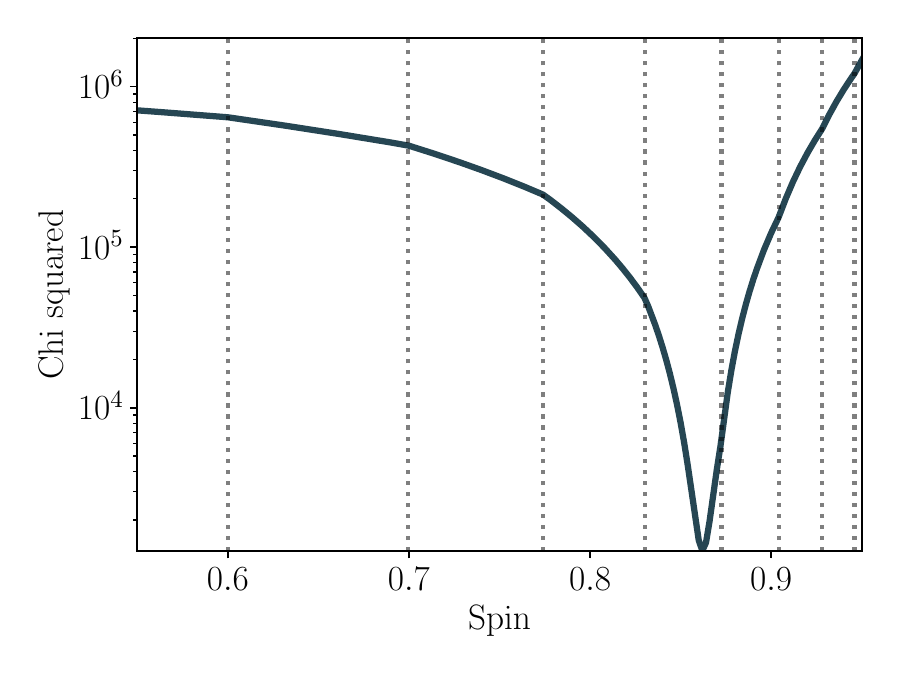}
    \includegraphics[width=1\columnwidth]{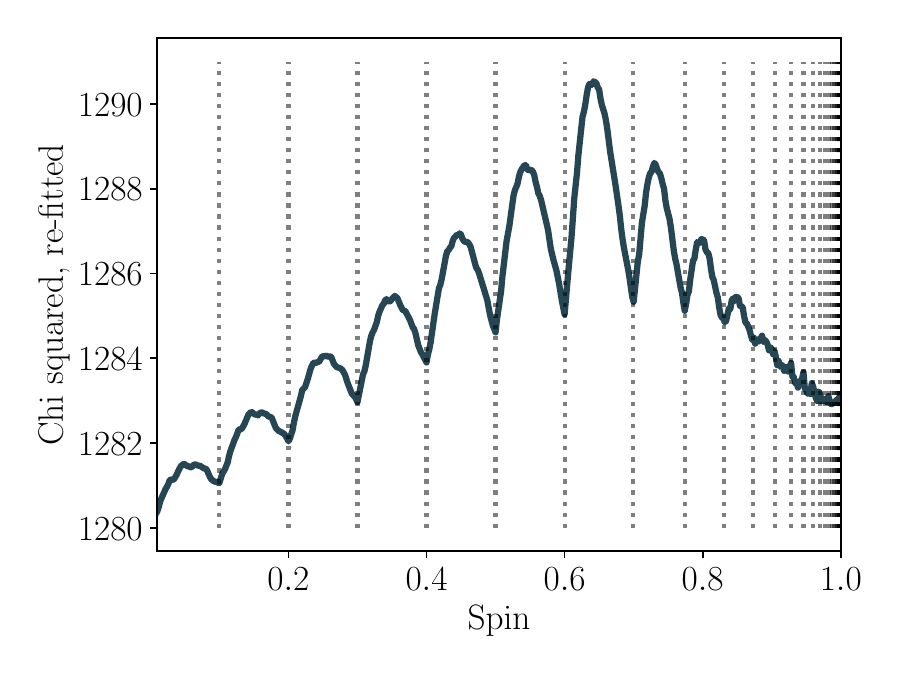}
    \caption{Top panel: fit statistic as a function of black hole spin. As in the top panel of Fig.\ref{fig:pl_fitstats}, the likelihood gradient is discontinuous.}
    \label{fig:spin_statistic}
\end{figure}

The shape of this distribution is thus very heavily warped by the linear interpolation. We can further demonstrate this by calculating the fit statistic as a function of spin, leaving all the other parameters fixed to the posterior median (top plot in Fig.\ref{fig:spin_statistic}), and re-fitting only the parameters of \texttt{kerrbb} while freezing the others for simplicity (bottom plot in Fig.\ref{fig:spin_statistic}). In the top panel we see that identically to the plot in Fig.\ref{fig:pl_fitstats}, the slope of the fit statistic changes discretely near the grid points. The bottom panel, which replicates that in the Appendix A of \cite{Parker19}, shows instead that the discrete jumps in the posterior reflect sharp increases in the likelihood away from the grid points (note that the fit statistic seems to improve for lower values of the spin, but it is unclear whether this would be the case if all the model parameters were left free). As \cite{Parker19} show, this behaviour can be mitigated with a non-linear interpolation scheme. In light of these numerical issues, it is difficult to state what the ``correct'' posterior of the spin would be. We note in addition that in order to produce this estimate we did not account for the uncertainty on the black hole mass, nor a potential contribution of the plunging region, which will introduce additional degeneracy and broaden the posterior distribution further \citep{Mummery25}, thus leading potentially to even more complex posterior distributions.

Finally, we show that the presence of a single parameter with a complex, multimodal posterior (i.e.~a posterior that MCMC samplers like \texttt{emcee} struggle to reconstruct), as is the case for the spin here, affects the behaviour of the entire ensemble, and therefore the inference of all the other parameters. We do this by comparing the auto-correlation lengths $\tau$ and posterior distributions as a chain is running, using the normalisation of the black-body component as an example. This model parameter is entirely un-correlated with the black hole spin (Fig.\ref{fig:posterior_freespin} in the Appendix), so one might expect that sampling its posterior would not be affected by the spin being free or fixed. This is, however, not the case.

\begin{figure*}
    \centering
    \includegraphics[width=\textwidth]{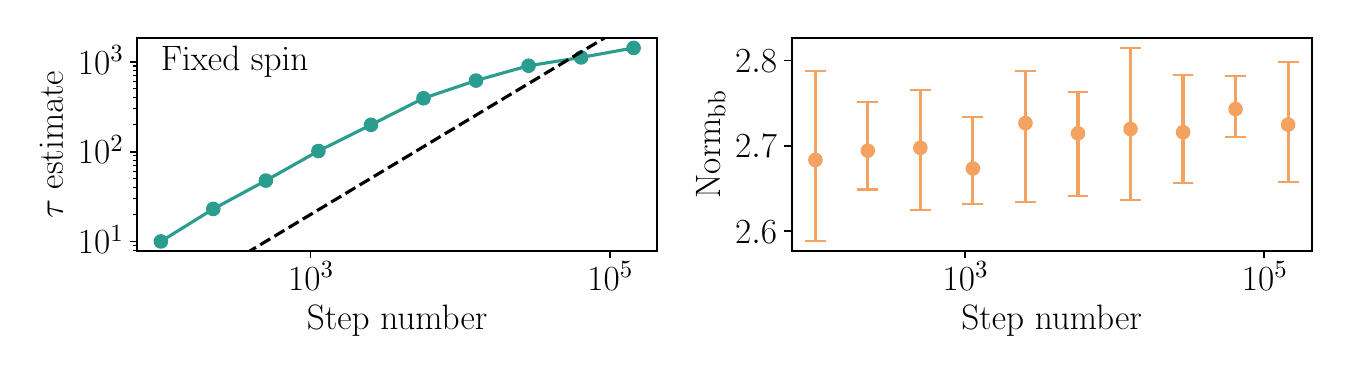}
    \includegraphics[width=\textwidth]{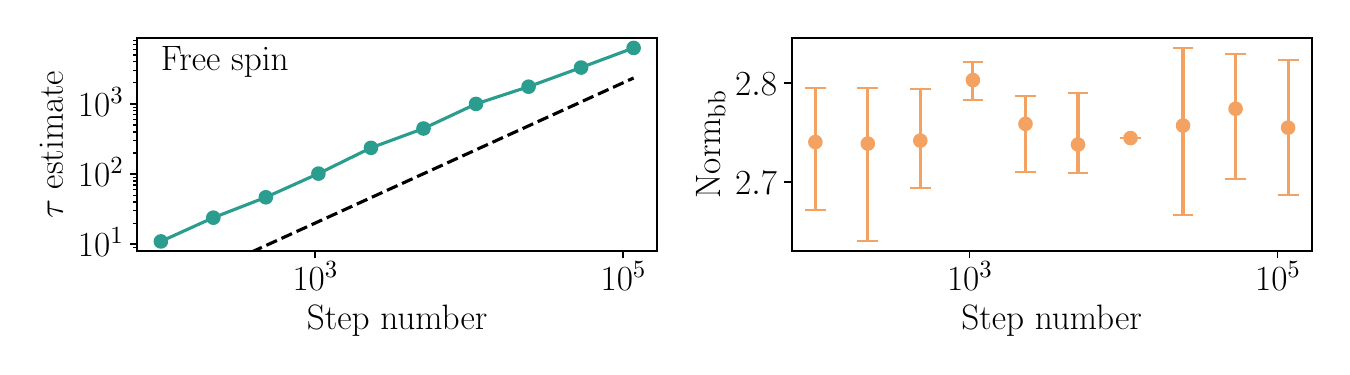}
    \caption{Estimated auto-correlation lengths (left column) and mean/credible range (right column) as a function of chain length. In the left column, the point of convergence (defined as number of steps $>50\tau$) is marked by the dashed line. The top row shows the (converged) run with fixed spin; the bottom row shows the (not converged) run with free spin. Note how in the first fit (top row), $\tau$ is far smaller and the credible range is more consistent across the chain, compared to the second fit (bottom row).}
    \label{fig:convergence}
\end{figure*}

We replicate the convergence analysis from the \texttt{emcee} documentation page\footnote{https://emcee.readthedocs.io/en/stable/tutorials/autocorr/}, and create a logarithmically spaced set of steps $N$ from 100 to the full 150000 steps that make up the chain. We then take the full chain up to $N$ and compute $\tau$. After discarding the burn-in and thinning the chain as before, we compute the 16th, 50th, and 84th percentile intervals for each value of $N$ as proxies for how the posterior estimates are evolving over the full run of the chain. These estimates are plotted in Fig.\ref{fig:convergence}, in which the difference between the top and bottom rows is in the spin being fixed/free, respectively. We see that in the first case (top row), $\tau$ evens out at $\approx1500$ - this means that our chain has been run long enough that we have sampled the longest variability scales in the path of the walkers. Because the chain ran longer than $N/50$ (shown in the black dashed line), following \cite{Hogg18} we conclude that the chain has converged to a reasonable estimate of  the posterior (and of $\tau$). In the second case (bottom row) the estimate for $\tau$ keeps increasing without stabilising, indicating that the values here only constitute a lower limit. In addition, the estimates of $\tau$ are always larger than $N/50$: this indicates that the second sampler likely did not yet sample the full posterior and therefore has not yet converged. Without running the MCMC chains for much longer, we can not be certain we reconstructed the posterior. Indeed, we note that the credible intervals in the posterior estimates as a function of $N$ (right column) are consistent across the chain only for the converged sampler. 

Finally, we also note that in order to converge we had to run the sampler for a rather long time. Using $\approx 10^{2}$ total walkers for $\approx 10^{5}$ steps results in $\approx 10^{7}$ model evaluations, or $\approx$ a day if each evaluation takes on the order of tens of milliseconds. Crucially, in \texttt{Xspec} users set the number of total likelihood evaluations, not the number of AIES steps, which may provide users with a false sense of confidence. Shorter sampling runs than these would incur a similar problem, and not sample the posterior distribution sufficiently. Such long inference times mean that robustly modelling many observations simultaneously, and deriving meaningful constraints from all of them, demands extreme amounts of computational power, if it is possible at all with standard software and techniques.

\subsection{Reflection modelling of a BHXRB hard state}
\label{sec:hardstate}

In the previous section, we discussed the impact a single, multi-modal parameter has on the inferred posteriors through MCMC sampling. In this section, we discuss an example in which  the multi-modality induced by using linear interpolation combines with large degeneracies among multiple parameters. Multi-modality and degeneracy combine to create a parameter space that is incredibly challenging to sample fully. As an illustrative example, we use reflection modelling in the hard state.

\begin{table}[]
    \centering
    \footnotesize{
    \begin{tabular}{c|c}
        Parameter & Prior  \\ \hline
        nH ($cm^{-2}$) & Uniform $[0.05,0.5$] \\ \hline
        Inclination (deg) & Normal$^{*}$ mean=$65^{\circ}$, $\sigma=4^{\circ}$ \\  
        $R_{\rm in}$ (R$_g$) & Uniform $[1,40]$ \\
        Emissivity index $q$ & Normal$^{**}$ mean=$3$, $\sigma=1$ \\
        Photon index $\Gamma$ & Uniform $[1.1,2.6]$ \\
        $\log(\xi)$ & Uniform $[0,4.7]$ \\ 
        $\log(N)$ & Uniform $[15,20]$ \\ 
        $A_{\rm Fe}$ & Uniform $[0.5,10]$ \\ 
        $kT_{\rm e}$ (keV) & Uniform $[20,400]$ \\
        Reflection fraction $f_{\rm ref}$ & Uniform $[0.5,5]$ \\ 
        \texttt{RelxillDCp} normalisation & Uniform $[0.01,1]$ \\ 
        \texttt{XillverDCp} normalisation$^{***}$ & Uniform $[0.001,0.1]$ \\ \hline 
        FPMB cross-calibration & Uniform $[0.75,1.25]$ \\ \hline
    \end{tabular}
    }
    \newline
    $^{*}$ From \cite{Atri20}
    \newline
    $^{**}$ Mimicking a horizontally extended corona as implied by IXPE observations \citep{Ixpe1}
    \newline
     $^{***}$ The photon index, temperature, Iron abundance and inclination in \texttt{XillverDCp} are tied to that of \texttt{RelxillDCp}, while the ionisation and density are fixed to their default values $\log(\xi)=0$, $\log(N)=15$ cm$^{-3}$
    \caption{Model parameters and priors used to fit the hard state data}
    \label{tab:relxill}
\end{table}

In this example, we use a typical setup consisting of both relativistic reflection (using the \texttt{RelxillDCp} model) and distant reflection (using \texttt{XillverDCp}), and including an absorption component with \texttt{tbabs}. We first attempted to use the standard methodology from the previous section and used in most analysis of BHXRBs: find an initial guess for the AIES sampler through an optimizer, use that guess to initialize the sampler, and evolve the MCMC walkers from there. The list of model free parameters and their priors is in Tab.\ref{tab:relxill}.

Once again, optimizing the fit statistic gave a reasonable fit, with $\chi^{2}/\rm{d.o.f.}\approx1.05$. However, we found that similarly to the previous section when including the spin as a parameter, the sampler behaved poorly, struggled to converge, and generally returned extremely large auto-correlation times ($\tau\gtrapprox$ thousands). Using more advanced MCMC like the ZEUS Slice Ensemble Sampler \citep{zeus} only partially improved the situation. In addition, we found that the posteriors were highly sensitive to the initial conditions, and often would shift significantly in different runs. This behaviour indicates that the parameter space is multi-modal, and crucially a single run is highly unlikely to recover all the modes successfully.

Instead, we used dynamic nested sampling using \texttt{Dynesty} \citep{Dynesty}. Nested sampling starts from the full prior volume and shrinks the available space along surfaces of improving likelihood. The initial conditions are limited to the choice of priors and the number of live points, and the latter can have a significant impact on the accuracy of the posteriors. Dynamic nested sampling allows users to favour crossing the prior volume quickly in order to estimate the posterior accurately, or to allocate more computational power to exploring the prior volume fully. The latter is required in order to identify multiple modes in the posterior distribution and can produce a more accurate estimate of the evidence. Due to the complexity of this particular problem, we found we had to use the latter. We achieved this by setting the weighting parameter to $f_p=0.5$, and the stopping parameter $s_p = 0.5$. For more information on the algorithm and its hyper-parameters, we refer the readers to \cite{Dynesty}. Our choice resulted in the sampler starting with 2800 live points, increasing up to $\approx$ 5000 before the run stopped. Using a more standard setup with 400-800 live points resulted in the sampler identifying only one of the posterior modes. 

With these settings and for this problem, \texttt{Dynesty} is dramatically more efficient than \texttt{emcee}. We can quantify this by comparing the effective sample size (ESS) produced by the two algorithms. For \texttt{emcee}, we ran the sampler for 180000 steps, using 130 walkers; this results in $\approx23$ million model evaluations and a runtime of $\approx30-40$ hours on a 12-core laptop CPU. Estimating $\tau$ at the end of the run returns $\approx5000-10000$, indicating that the sampler failed to converge; regardless, we can optimistically estimate the ESS to be $n_{\rm steps}\times n_{\rm walkers}/\tau\approx 2300$. On the other hand, \texttt{Dynesty} allows the ESS to be estimated directly from the results of a run; the built-in functions return ESS$\approx 50000$ for $\approx$ 5 million likelihood evaluations, which on the same CPU took 10 hours. Between the larger ESS and shorter run-time, in this particular case \texttt{Dynesty} performs at least $\approx$ 80 times better than \texttt{emcee}, and therefore we only report results for the former. We note that with such a large ESS we can be somewhat confident that \texttt{Dynesty} reconstructed the posteriors reasonably accurately, at least for the modes it is able to identify (which we discuss later). Finally, we also note that this comparison is somewhat unfair, as \texttt{emcee} is known for being ill-suited to highly complex problems, but it serves as an illustration that different sampling problems are better suited to, and should be tackled with, different sampling algorithms. 

\begin{figure}
    \centering
    \includegraphics[width=\columnwidth]{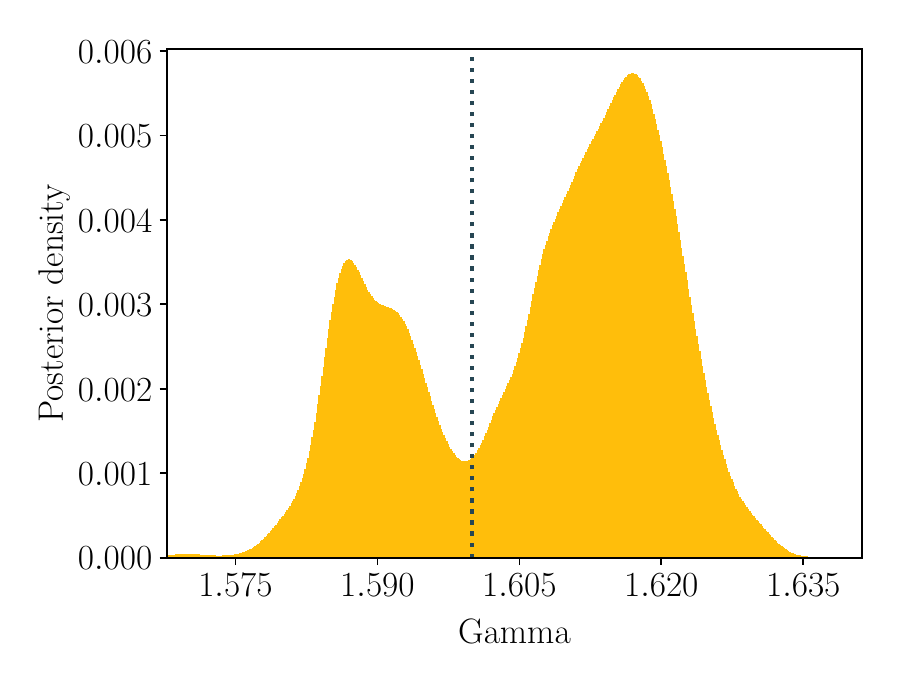}
    \caption{Posterior distribution of the photon index. The posterior is bimodal, with the two modes separated exactly by the model grid point (dashed line). }
    \label{fig:posterior_gamma}
\end{figure}

\begin{figure}
    \centering
    \includegraphics[width=\columnwidth]{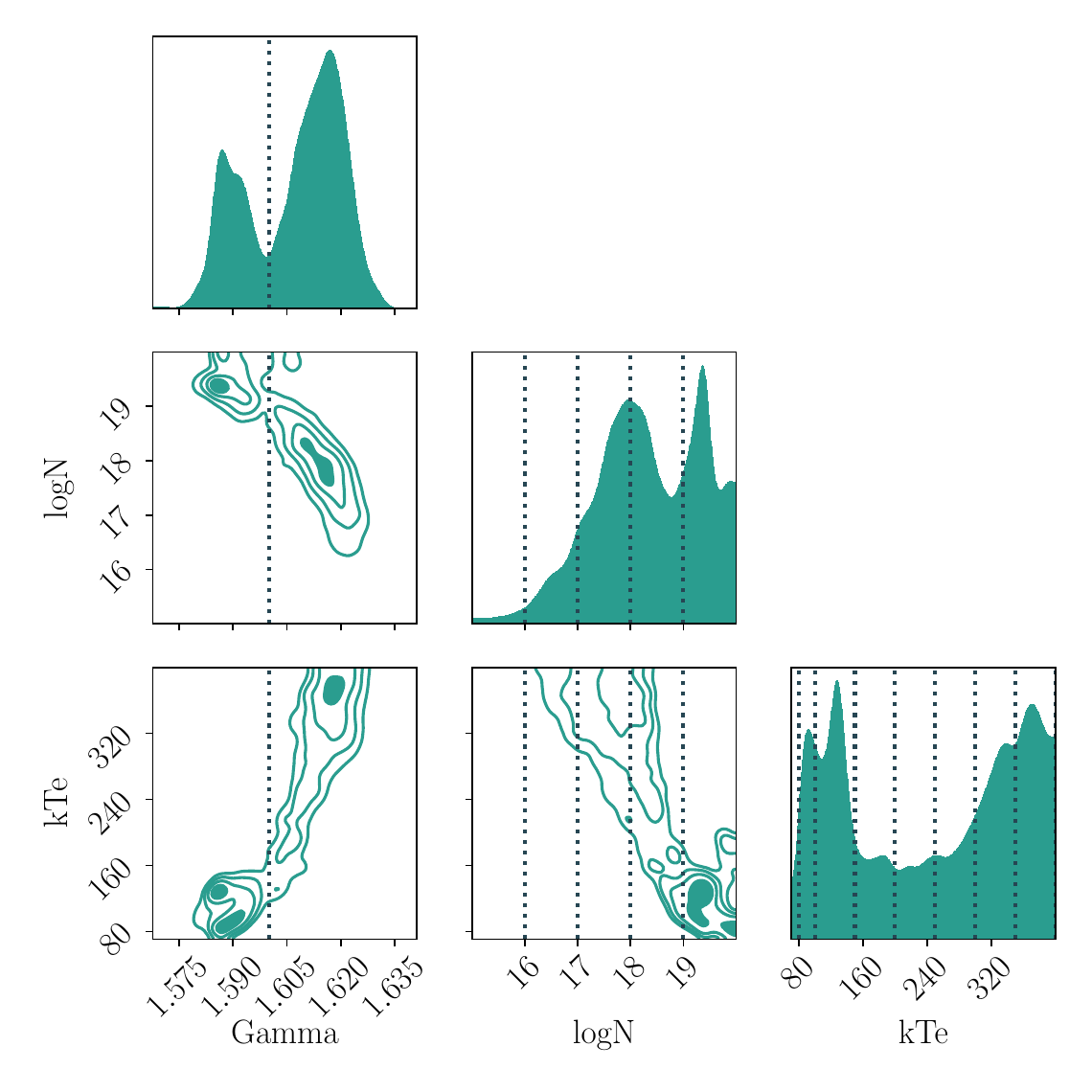}
     \caption{Two-dimensional posteriors of the photon index $\Gamma$, reflector density $\log(N)$, and temperature $kT_{\rm e}$. Note how a) the bi-modality in $\Gamma$ propagates to other parameters, and b) $\log(N)$ and $kT_{\rm e}$ themselves show features somewhat correlated with the grid points (marked by the dashed lines).}
    \label{fig:posterior_gamma_2d}
\end{figure}

Fig.\ref{fig:posterior_gamma} shows the one-dimensional posterior for the photon index $\Gamma$. We see clearly that there are two peaks, separated exactly where a \texttt{Xillver} grid point ($\Gamma=1.6$) lies. This separation is consistent with the regime discussed in Sec.\ref{sec:regime_two}, and is therefore likely caused by linear interpolation. Fig.\ref{fig:posterior_gamma_2d} shows a corner plot, including only the parameters whose posteriors are affected by this artificial feature: in this case, the temperature and density. We clearly see that a) the temperature also shows a bimodal distribution, caused by the correlation between it and $\Gamma$ and b) the density and temperature show additional features that are (somewhat) correlated with the grid points for these parameters. This suggests that the posteriors are also affected by the third regime discussed in Sec.\ref{sec:regime_three}; the combination of this multi-modality, with the correlations between parameters shown here, means that this parameter space is effectively impossible to sample with standard optimizers or AIES samplers used in most analysis. In turn, this difficulty in sampling might partially explain why reflection results tend to be so inconsistent and/or different authors often find different solutions \citep[e.g.][]{Connors22}: as we have shown in Sec.\ref{sec:trivial} and Sec.\ref{sec:powerlaw}, the behaviour is very heavily driven by the linear interpolation used, and may not represent a real feature of the model, nor of real accretion physics. 
%It is entirely self fucking inflicted fuck you fucking idiots. 

\section{Discussion and conclusion}
\label{sec:Discussion}

We have demonstrated that the use of linear interpolation to interpolate non-linear table models, even as simple as a power-law, can severely compromise the ability to reconstruct accurate parameter posteriors distributions. While this behaviour is well established in applied mathematics and statistics, here we have demonstrated that it has a large impact on the ability of some X-ray spectral models to derive meaningful constraints from fits to modern, high quality data. In particular, the following issues can arise:

\begin{itemize}
    \item If the model posteriors are narrow enough to not cross multiple grid points, then the farther the true posteriors are from a grid point, the more biased the estimate of the parameter will be (Sec.\ref{sec:trivial} and \ref{sec:regime_one}). This bias is progressively larger for coarser model grids.
    \item When the model posterior crosses a grid point, particularly if the model is misspecified, the posterior distribution can become bi-modal, with a valley or peak in the posterior appearing in correspondence of each grid point (Sec.\ref{sec:trivial} and \ref{sec:regime_two}).
    \item When the model posteriors are broad (for example, because of degeneracy) and cross over multiple grid points, the recovered posteriors can become artificially multi-modal, with peaks in the posterior corresponding to each grid point. Depending on the grid and parameters, this multi-modality can be severe enough to cause a catastrophic failure in the performance of standard samplers (Sec.\ref{sec:trivial}, \ref{sec:regime_three}, and \ref{sec:softstate}).
    \item If parameters display both multi-modal distributions and are degenerate with other parameters, then the multi-modality can propagate throughout the parameter space, resulting in additional features and further degrading sampler performance (Sec.\ref{sec:hardstate}).  
\end{itemize}

Due to these issues, we find that even advanced samplers like \texttt{Dynesty} can lead users into a false sense of confidence: we only succeeded in identifying the multi-modal spaces described here using non-standard sampler settings that greatly increase the run-time and computational cost of a single fit. On the other hand, setting up the sampler with more standard hyper-parameters, like one would expect from a reasonably well-behaved parameter space, caused the run to converge very well to a single solution. 

While we have discussed our findings within a Bayesian framework, Frequentist approaches \citep[e.g.][]{Shafee06,McClintock06,Risaliti13,Garcia15} are affected just as badly. Frequentist model selection is performed by comparing changes in the fit statistic, which loses meaning when some of those changes might arise from the computational artifacts shown in Sec.\ref{sec:trivial}. In addition, software like \texttt{Xspec} and \texttt{ISIS} by default use the Levenberg-Marquardt algorithm, which is a local optimizer that tries to estimate the gradient of the likelihood to search for the best-fitting solution and compute the uncertainties through the \texttt{error} and \texttt{steppar} commands. Here we have demonstrated that the gradient of the likelihood is discontinuous, which causes the gradient information to lose meaning. Indeed, \texttt{Xspec} documentation recommends users switch to the \texttt{migrad}/\texttt{minuit} algorithms when fitting table models.% Throughout the literature it is rare to find mentions of the exact algorithm set for a given \texttt{Xspec} fit; however, it is likely that some of the results published may have used incorrect settings for the problem, leading to incorrect estimates of either the best-fitting parameters, or their uncertainties.

We note that linear interpolation of table models is not necessarily the only cause for complex and/or biased posteriors. For example, model misspecification can play a similar role, and indeed \cite{Parker22} demonstrate that it can also result in highly biased and/or overly confident estimates of fundamental parameters like black hole spin. We can not demonstrate that model misspecification does not play a role in shaping the posteriors we derive in Sec.\ref{sec:RealObservations}; however, compared to the AGN discussed in \cite{Parker22}, the observations we analysed here should be relatively clean, because they cover only hard X-rays, and because the source itself is a standard, well-behaved, and extremely bright source. A more in-depth treatment of model misspecification, and statistical techniques for mitigating it, will be covered in a future paper (Ricketts et al. in prep).

Given the increasing quality of data available to X-ray astronomers, these are issues which must be addressed before next-generation missions like \textit{eXTP} and \textit{NewAthena} begin observing. While such fundamental problems can not easily be solved in the short to medium term, they can be mitigated. 

First, more care should be taken in verifying and reporting the convergence of a given sampler run or optimiser quantitatively; in general, we believe that the community should run their samplers for far more steps than is done currently. As we discussed in the previous sections, users should set the burn-in length of a chain to be on the order of 2 autocorrelation lengths $\tau$, and to run the AIES sampler for $\approx 50\tau$ as a reasonable standard to minimize Monte Carlo errors in sampling.  Yet, it is common for the lengths of the chains reported in literature to be only a few times the burn-in, suggesting that posterior distributions are often undersampled. 

Second, whenever possible, scientists should verify their fit results by comparing tabulated models with on-the-fly computations, as a simple convergence check that the physics retrieved in both computational approaches are consistent. This is possible, for example, for non-equilibrium plasma models in the \texttt{Spex} package \citep{spex}, but is not a universal solution. 

Third, the community should strive to consistently provide access to reproduction packages linked to each publication, including end-to-end scripts capable of replicating each fit as reported; the inability to reproduce a given fit would immediately hint at the presence of the numerical issues discussed here. 

Finally, we encourage the community to explore different sampling algorithms in their research. There are multiple variants of both MCMC and nested sampling algorithms, each with strengths and weaknesses, and given the (often artificial) complexity of the parameter spaces discussed here, more exploration of advanced sampler techniques is required. A handful of advanced samplers are available in \texttt{Xspec} as of version 13.0, and more algorithms can be accessed through alternative packages like \texttt{BXA} (Bayesian X-ray Analysis, \citealt{bxa}), \texttt{Jaxspec} \citep{jaxspec} or \texttt{nDspec} \citep{Lucchini26}. In particular, we caution users to steer away from any gradient-based method, either during optimisation or posterior sampling, when using a table model.

Regardless of short term mitigation strategies, the issues highlighted here are fundamental computational problems that can only be solved with better software; options include using more complex interpolation like higher order polynomials \citep{Parker19}, Gaussian process regression \citep{Foreman-Mackey17}, or surrogate models \citep{jaxspec,Ricketts25}. We plan to include a more advanced interpolation package in \texttt{nDspec} in the near future (Baker et al. in prep.), which we expect will reduce interpolation errors considerably. 

\begin{acknowledgements}
%ML thanks certain senior astronomers for greatly helping in his future career choices.
GM acknowledges support from the Academy of Finland grant 355672. This paper made use of code from \textsc{Astropy} (\citealt{astropy13}, \citealt{astropy18}, \citealt{astropy22}), \textsc{NumPy} \citep{numpy}, \textsc{SciPy} \citep{scipy}, \textsc{LMFIT} \citep{lmfit}, \textsc{matplotlib} \citep{matplotlib}, \textsc{ndspec} \citep{Lucchini26}, \textsc{emcee} \citep{emcee}, \textsc{Dynesty} \citep{Dynesty} and \textsc{zeus} \citep{zeus}. 
\end{acknowledgements}

\bibliography{references}

@ARTICLE{Albert25,
       author = {{Albert}, Joshua and {Bal{\'a}zs}, Csaba and {Fowlie}, Andrew and {Handley}, Will and {Hunt-Smith}, Nicholas and {Ruiz de Austri}, Roberto and {White}, Martin},
        title = "{A comparison of Bayesian sampling algorithms for high-dimensional particle physics and cosmology applications}",
      journal = {Computer Physics Communications},
         year = 2025,
        month = oct,
       volume = {315},
          eid = {109756},
        pages = {109756},
          doi = {10.1016/j.cpc.2025.109756},
archivePrefix = {arXiv},
       eprint = {2409.18464},
 primaryClass = {hep-ph},
       adsurl = {https://ui.adsabs.harvard.edu/abs/2025CoPhC.31509756A}
}

@ARTICLE{Fabian20,
       author = {{Fabian}, A.~C. and {Buisson}, D.~J. and {Kosec}, P. and {Reynolds}, C.~S. and {Wilkins}, D.~R. and {Tomsick}, J.~A. and {Walton}, D.~J. and {Gandhi}, P. and {Altamirano}, D. and {Arzoumanian}, Z. and {Cackett}, E.~M. and {Dyda}, S. and {Garcia}, J.~A. and {Gendreau}, K.~C. and {Grefenstette}, B.~W. and {Homan}, J. and {Kara}, E. and {Ludlam}, R.~M. and {Miller}, J.~M. and {Steiner}, J.~F.},
        title = "{The soft state of the black hole transient source MAXI J1820+070: emission from the edge of the plunge region?}",
      journal = {\mnras},
         year = 2020,
        month = apr,
       volume = {493},
       number = {4},
        pages = {5389-5396},
          doi = {10.1093/mnras/staa564},
archivePrefix = {arXiv},
       eprint = {2002.09691},
 primaryClass = {astro-ph.HE},
       adsurl = {https://ui.adsabs.harvard.edu/abs/2020MNRAS.493.5389F}
}

@INPROCEEDINGS{Skilling04,
       author = {{Skilling}, John},
        title = "{Nested Sampling}",
    booktitle = {Bayesian Inference and Maximum Entropy Methods in Science and Engineering: 24th International Workshop on Bayesian Inference and Maximum Entropy Methods in Science and Engineering},
         year = 2004,
       editor = {{Fischer}, Rainer and {Preuss}, Roland and {Toussaint}, Udo Von},
       series = {American Institute of Physics Conference Series},
       volume = {735},
        month = nov,
    publisher = {AIP},
        pages = {395-405},
          doi = {10.1063/1.1835238},
       adsurl = {https://ui.adsabs.harvard.edu/abs/2004AIPC..735..395S}
}

@ARTICLE{GoodmanWeare2010,
       author = {{Goodman}, Jonathan and {Weare}, Jonathan},
        title = "{Ensemble samplers with affine invariance}",
      journal = {Communications in Applied Mathematics and Computational Science},
         year = 2010,
        month = jan,
       volume = {5},
       number = {1},
        pages = {65-80},
          doi = {10.2140/camcos.2010.5.65},
       adsurl = {https://ui.adsabs.harvard.edu/abs/2010CAMCS...5...65G}
}

@ARTICLE{Parker19,
       author = {{Parker}, M.~L. and {Buisson}, D.~J.~K. and {Tomsick}, J.~A. and {Fabian}, A.~C. and {Madsen}, K.~K. and {Walton}, D.~J. and {F{\"u}rst}, F.},
        title = "{XRB continuum fitting with sensitive high-energy X-ray detectors}",
      journal = {\mnras},
         year = 2019,
        month = mar,
       volume = {484},
       number = {1},
        pages = {1202-1212},
          doi = {10.1093/mnras/stz045},
archivePrefix = {arXiv},
       eprint = {1901.00683},
 primaryClass = {astro-ph.HE},
       adsurl = {https://ui.adsabs.harvard.edu/abs/2019MNRAS.484.1202P}
}

@ARTICLE{Hogg18,
       author = {{Hogg}, David W. and {Foreman-Mackey}, Daniel},
        title = "{Data Analysis Recipes: Using Markov Chain Monte Carlo}",
      journal = {\apjs},
         year = 2018,
        month = may,
       volume = {236},
       number = {1},
          eid = {11},
        pages = {11},
          doi = {10.3847/1538-4365/aab76e},
archivePrefix = {arXiv},
       eprint = {1710.06068},
 primaryClass = {astro-ph.IM},
       adsurl = {https://ui.adsabs.harvard.edu/abs/2018ApJS..236...11H}
}

@article{Kennedy01,
  author  = {Kennedy, Marc C. and O'Hagan, Anthony},
  title   = {Bayesian calibration of computer models},
  journal = {Journal of the Royal Statistical Society: Series B (Statistical Methodology)},
  year    = {2001},
  volume  = {63},
  number  = {3},
  pages   = {425--464},
  doi     = {10.1111/1467-9868.00294}
}

@ARTICLE{reflionx,
       author = {{Ross}, R.~R. and {Fabian}, A.~C. and {Young}, A.~J.},
        title = "{X-ray reflection spectra from ionized slabs}",
      journal = {\mnras},
         year = 1999,
        month = jun,
       volume = {306},
       number = {2},
        pages = {461-466},
          doi = {10.1046/j.1365-8711.1999.02528.x},
archivePrefix = {arXiv},
       eprint = {astro-ph/9902325},
 primaryClass = {astro-ph},
       adsurl = {https://ui.adsabs.harvard.edu/abs/1999MNRAS.306..461R}
}

@ARTICLE{Davis06,
       author = {{Davis}, Shane W. and {Hubeny}, Ivan},
        title = "{A Grid of Relativistic, Non-LTE Accretion Disk Models for Spectral Fitting of Black Hole Binaries}",
      journal = {\apjs},
         year = 2006,
        month = jun,
       volume = {164},
       number = {2},
        pages = {530-535},
          doi = {10.1086/503549},
archivePrefix = {arXiv},
       eprint = {astro-ph/0602499},
 primaryClass = {astro-ph},
       adsurl = {https://ui.adsabs.harvard.edu/abs/2006ApJS..164..530D}
}

@article{Kaipio07,
  author  = {Kaipio, Jari and Somersalo, Erkki},
  title   = {Statistical inverse problems: {D}iscretization, model reduction and inverse crimes},
  journal = {Journal of Computational and Applied Mathematics},
  year    = {2007},
  volume  = {198},
  number  = {2},
  pages   = {493--504},
  doi     = {10.1016/j.cam.2005.09.027}
}

@article{Bates80,
  author  = {Bates, Douglas M. and Watts, Donald G.},
  title   = {Relative Curvature Measures of Nonlinearity},
  journal = {Journal of the Royal Statistical Society: Series B (Methodological)},
  year    = {1980},
  volume  = {42},
  number  = {1},
  pages   = {1--25},
  doi     = {10.1111/j.2517-6161.1980.tb01094.x}
}

@ARTICLE{Li05,
       author = {{Li}, Li-Xin and {Zimmerman}, Erik R. and {Narayan}, Ramesh and {McClintock}, Jeffrey E.},
        title = "{Multitemperature Blackbody Spectrum of a Thin Accretion Disk around a Kerr Black Hole: Model Computations and Comparison with Observations}",
      journal = {\apjs},
         year = 2005,
        month = apr,
       volume = {157},
       number = {2},
        pages = {335-370},
          doi = {10.1086/428089},
archivePrefix = {arXiv},
       eprint = {astro-ph/0411583},
 primaryClass = {astro-ph},
       adsurl = {https://ui.adsabs.harvard.edu/abs/2005ApJS..157..335L}
}

@ARTICLE{Foreman-Mackey17,
       author = {{Foreman-Mackey}, Daniel and {Agol}, Eric and {Ambikasaran}, Sivaram and {Angus}, Ruth},
        title = "{Fast and Scalable Gaussian Process Modeling with Applications to Astronomical Time Series}",
      journal = {\aj},
         year = 2017,
        month = dec,
       volume = {154},
       number = {6},
          eid = {220},
        pages = {220},
          doi = {10.3847/1538-3881/aa9332},
archivePrefix = {arXiv},
       eprint = {1703.09710},
 primaryClass = {astro-ph.IM},
       adsurl = {https://ui.adsabs.harvard.edu/abs/2017AJ....154..220F}
}

@ARTICLE{Parker22,
       author = {{Parker}, M.~L. and {Matzeu}, G.~A. and {Matthews}, J.~H. and {Middleton}, M.~J. and {Dauser}, T. and {Jiang}, J. and {Joyce}, A.~M.},
        title = "{The X-ray disc/wind degeneracy in AGN}",
      journal = {\mnras},
         year = 2022,
        month = jun,
       volume = {513},
       number = {1},
        pages = {551-572},
          doi = {10.1093/mnras/stac877},
archivePrefix = {arXiv},
       eprint = {2203.14789},
 primaryClass = {astro-ph.HE},
       adsurl = {https://ui.adsabs.harvard.edu/abs/2022MNRAS.513..551P}
}

@ARTICLE{Mummery25,
       author = {{Mummery}, Andrew and {Jiang}, Jiachen and {Ingram}, Adam and {Fabian}, Andrew and {Rule}, Jake},
        title = "{A rapid black hole spin or emission from the plunging region?}",
      journal = {\mnras},
         year = 2025,
        month = dec,
       volume = {544},
       number = {3},
        pages = {2880-2896},
          doi = {10.1093/mnras/staf1758},
archivePrefix = {arXiv},
       eprint = {2505.13119},
 primaryClass = {astro-ph.HE},
       adsurl = {https://ui.adsabs.harvard.edu/abs/2025MNRAS.544.2880M}
}

@ARTICLE{Risaliti13,
       author = {{Risaliti}, G. and {Harrison}, F.~A. and {Madsen}, K.~K. and {Walton}, D.~J. and {Boggs}, S.~E. and {Christensen}, F.~E. and {Craig}, W.~W. and {Grefenstette}, B.~W. and {Hailey}, C.~J. and {Nardini}, E. and {Stern}, Daniel and {Zhang}, W.~W.},
        title = "{A rapidly spinning supermassive black hole at the centre of NGC 1365}",
      journal = {\nat},
         year = 2013,
        month = feb,
       volume = {494},
       number = {7438},
        pages = {449-451},
          doi = {10.1038/nature11938},
archivePrefix = {arXiv},
       eprint = {1302.7002},
 primaryClass = {astro-ph.HE},
       adsurl = {https://ui.adsabs.harvard.edu/abs/2013Natur.494..449R}
}

@ARTICLE{McClintock06,
       author = {{McClintock}, Jeffrey E. and {Shafee}, Rebecca and {Narayan}, Ramesh and {Remillard}, Ronald A. and {Davis}, Shane W. and {Li}, Li-Xin},
        title = "{The Spin of the Near-Extreme Kerr Black Hole GRS 1915+105}",
      journal = {\apj},
         year = 2006,
        month = nov,
       volume = {652},
       number = {1},
        pages = {518-539},
          doi = {10.1086/508457},
archivePrefix = {arXiv},
       eprint = {astro-ph/0606076},
 primaryClass = {astro-ph},
       adsurl = {https://ui.adsabs.harvard.edu/abs/2006ApJ...652..518M}
}

@ARTICLE{Shafee06,
       author = {{Shafee}, Rebecca and {McClintock}, Jeffrey E. and {Narayan}, Ramesh and {Davis}, Shane W. and {Li}, Li-Xin and {Remillard}, Ronald A.},
        title = "{Estimating the Spin of Stellar-Mass Black Holes by Spectral Fitting of the X-Ray Continuum}",
      journal = {\apjl},
         year = 2006,
        month = jan,
       volume = {636},
       number = {2},
        pages = {L113-L116},
          doi = {10.1086/498938},
archivePrefix = {arXiv},
       eprint = {astro-ph/0508302},
 primaryClass = {astro-ph},
       adsurl = {https://ui.adsabs.harvard.edu/abs/2006ApJ...636L.113S}
}

@ARTICLE{Connors22,
       author = {{Connors}, Riley M.~T. and {Garc{\'\i}a}, Javier A. and {Tomsick}, John and {Mastroserio}, Guglielmo and {Grinberg}, Victoria and {Steiner}, James F. and {Jiang}, Jiachen and {Fabian}, Andrew C. and {Parker}, Michael L. and {Harrison}, Fiona and {Hare}, Jeremy and {Mallick}, Labani and {Lazar}, Hadar},
        title = "{The Long-stable Hard State of XTE J1752-223 and the Disk Truncation Dilemma}",
      journal = {\apj},
         year = 2022,
        month = aug,
       volume = {935},
       number = {2},
          eid = {118},
        pages = {118},
          doi = {10.3847/1538-4357/ac7ff2},
archivePrefix = {arXiv},
       eprint = {2207.06567},
 primaryClass = {astro-ph.HE},
       adsurl = {https://ui.adsabs.harvard.edu/abs/2022ApJ...935..118C}
}

@ARTICLE{Lucchini26,
       author = {{Lucchini}, Matteo and {Ricketts}, Benjamin and {Uttley}, Phil and {Huppenkothen}, Daniela},
        title = "{nDspec: a new Python library for modelling multi-dimensional datasets in X-ray astronomy}",
      journal = {arXiv e-prints},
         year = 2025,
        month = dec,
          eid = {arXiv:2512.10615},
        pages = {arXiv:2512.10615},
          doi = {10.48550/arXiv.2512.10615},
archivePrefix = {arXiv},
       eprint = {2512.10615},
 primaryClass = {astro-ph.HE},
       adsurl = {https://ui.adsabs.harvard.edu/abs/2025arXiv251210615L}
}

@ARTICLE{Smith01,
       author = {{Smith}, Randall K. and {Brickhouse}, Nancy S. and {Liedahl}, Duane A. and {Raymond}, John C.},
        title = "{Collisional Plasma Models with APEC/APED: Emission-Line Diagnostics of Hydrogen-like and Helium-like Ions}",
      journal = {\apjl},
         year = 2001,
        month = aug,
       volume = {556},
       number = {2},
        pages = {L91-L95},
          doi = {10.1086/322992},
archivePrefix = {arXiv},
       eprint = {astro-ph/0106478},
 primaryClass = {astro-ph},
       adsurl = {https://ui.adsabs.harvard.edu/abs/2001ApJ...556L..91S}
}

@article{Roberts01,
author = {Gareth O. Roberts and Jeffrey S. Rosenthal},
title = {{Optimal scaling for various Metropolis-Hastings algorithms}},
volume = {16},
journal = {Statistical Science},
number = {4},
publisher = {Institute of Mathematical Statistics},
pages = {351 -- 367},
year = {2001},
doi = {10.1214/ss/1015346320},
URL = {https://doi.org/10.1214/ss/1015346320}
}

@ARTICLE{Ashton22,
       author = {{Ashton}, Greg and {Bernstein}, Noam and {Buchner}, Johannes and {Chen}, Xi and {Cs{\'a}nyi}, G{\'a}bor and {Fowlie}, Andrew and {Feroz}, Farhan and {Griffiths}, Matthew and {Handley}, Will and {Habeck}, Michael and {Higson}, Edward and {Hobson}, Michael and {Lasenby}, Anthony and {Parkinson}, David and {P{\'a}rtay}, Livia B. and {Pitkin}, Matthew and {Schneider}, Doris and {Speagle}, Joshua S. and {South}, Leah and {Veitch}, John and {Wacker}, Philipp and {Wales}, David J. and {Yallup}, David},
        title = "{Nested sampling for physical scientists}",
      journal = {Nature Reviews Methods Primers},
         year = 2022,
        month = may,
       volume = {2},
          eid = {39},
        pages = {39},
          doi = {10.1038/s43586-022-00121-x},
archivePrefix = {arXiv},
       eprint = {2205.15570},
 primaryClass = {stat.CO},
       adsurl = {https://ui.adsabs.harvard.edu/abs/2022NRvMP...2...39A}
}

@ARTICLE{zeus,
       author = {{Karamanis}, Minas and {Beutler}, Florian and {Peacock}, John A.},
        title = "{zeus: a PYTHON implementation of ensemble slice sampling for efficient Bayesian parameter inference}",
      journal = {\mnras},
         year = 2021,
        month = dec,
       volume = {508},
       number = {3},
        pages = {3589-3603},
          doi = {10.1093/mnras/stab2867},
archivePrefix = {arXiv},
       eprint = {2105.03468},
 primaryClass = {astro-ph.IM},
       adsurl = {https://ui.adsabs.harvard.edu/abs/2021MNRAS.508.3589K}
}

@ARTICLE{Dynesty,
       author = {{Speagle}, Joshua S.},
        title = "{DYNESTY: a dynamic nested sampling package for estimating Bayesian posteriors and evidences}",
      journal = {\mnras},
         year = 2020,
        month = apr,
       volume = {493},
       number = {3},
        pages = {3132-3158},
          doi = {10.1093/mnras/staa278},
archivePrefix = {arXiv},
       eprint = {1904.02180},
 primaryClass = {astro-ph.IM},
       adsurl = {https://ui.adsabs.harvard.edu/abs/2020MNRAS.493.3132S}
}

@ARTICLE{scipy,
  author  = {Virtanen, Pauli and Gommers, Ralf and Oliphant, Travis E. and
            Haberland, Matt and Reddy, Tyler and Cournapeau, David and
            Burovski, Evgeni and Peterson, Pearu and Weckesser, Warren and
            Bright, Jonathan and {van der Walt}, St{\'e}fan J. and
            Brett, Matthew and Wilson, Joshua and Millman, K. Jarrod and
            Mayorov, Nikolay and Nelson, Andrew R. J. and Jones, Eric and
            Kern, Robert and Larson, Eric and Carey, C J and
            Polat, {\.I}lhan and Feng, Yu and Moore, Eric W. and
            {VanderPlas}, Jake and Laxalde, Denis and Perktold, Josef and
            Cimrman, Robert and Henriksen, Ian and Quintero, E. A. and
            Harris, Charles R. and Archibald, Anne M. and
            Ribeiro, Ant{\^o}nio H. and Pedregosa, Fabian and
            {van Mulbregt}, Paul and {SciPy 1.0 Contributors}},
  title   = {{{SciPy} 1.0: Fundamental Algorithms for Scientific
            Computing in Python}},
  journal = {Nature Methods},
  year    = {2020},
  volume  = {17},
  pages   = {261--272},
  adsurl  = {https://rdcu.be/b08Wh},
  doi     = {10.1038/s41592-019-0686-2},
}

@software{spex,
  author       = {Kaastra, J. S. and
                  Raassen, A. J. J. and
                  de Plaa, J. and
                  Gu, Liyi},
  title        = {SPEX X-ray spectral fitting package},
  month        = oct,
  year         = 2025,
  publisher    = {Zenodo},
  version      = {3.08.02},
  doi          = {10.5281/zenodo.17313851},
  url          = {https://doi.org/10.5281/zenodo.17313851},
}

@ARTICLE{Ricketts25,
       author = {{Ricketts}, Benjamin J. and {Huppenkothen}, Daniela and {Lucchini}, Matteo and {Ingram}, Adam and {Mastroserio}, Guglielmo and {Ho}, Matthew and {Wandelt}, Benjamin},
        title = "{RTFAST-Spectra: emulation of X-ray reverberation mapping for active galactic nuclei}",
      journal = {\mnras},
         year = 2025,
        month = apr,
       volume = {538},
       number = {2},
        pages = {1096-1117},
          doi = {10.1093/mnras/staf337},
archivePrefix = {arXiv},
       eprint = {2412.10131},
 primaryClass = {astro-ph.HE},
       adsurl = {https://ui.adsabs.harvard.edu/abs/2025MNRAS.538.1096R}
}

@software{lmfit,
  author       = {Newville, Matthew and
                  Otten, Renee and
                  Nelson, Andrew and
                  Stensitzki, Till and
                  Ingargiola, Antonino and
                  Allan, Daniel and
                  Fox, Austin and
                  Carter, Faustin and
                  Rawlik, Michal},
  title        = {LMFIT: Non-Linear Least-Squares Minimization and
                   Curve-Fitting for Python
                  },
  month        = mar,
  year         = 2025,
  publisher    = {Zenodo},
  version      = {1.3.3},
  doi          = {10.5281/zenodo.15014437},
  url          = {https://doi.org/10.5281/zenodo.15014437},
  swhid        = {swh:1:dir:6dbda1361832412880d315cbf608ba498c177d82
                   ;origin=https://doi.org/10.5281/zenodo.598352;visi
                   t=swh:1:snp:0d236d4d8ff7f7248600297b145d047734607d
                   14;anchor=swh:1:rel:5f2a70f7c390d9b78cb661b349b044
                   753cf7e89e;path=lmfit-lmfit-py-f97ddf7
                  },
}

@ARTICLE{bxa,
       author = {{Buchner}, J. and {Georgakakis}, A. and {Nandra}, K. and {Hsu}, L. and {Rangel}, C. and {Brightman}, M. and {Merloni}, A. and {Salvato}, M. and {Donley}, J. and {Kocevski}, D.},
        title = "{X-ray spectral modelling of the AGN obscuring region in the CDFS: Bayesian model selection and catalogue}",
      journal = {\aap},
         year = 2014,
        month = apr,
       volume = {564},
          eid = {A125},
        pages = {A125},
          doi = {10.1051/0004-6361/201322971},
archivePrefix = {arXiv},
       eprint = {1402.0004},
 primaryClass = {astro-ph.HE},
       adsurl = {https://ui.adsabs.harvard.edu/abs/2014A&A...564A.125B}
}

@Article{matplotlib,
  Author    = {Hunter, J. D.},
  Title     = {Matplotlib: A 2D graphics environment},
  Journal   = {Computing in Science \& Engineering},
  Volume    = {9},
  Number    = {3},
  Pages     = {90--95},
  publisher = {IEEE COMPUTER SOC},
  doi       = {10.1109/MCSE.2007.55},
  year      = 2007
}

@Article{numpy,
 title         = {Array programming with {NumPy}},
 author        = {Charles R. Harris and K. Jarrod Millman and St{\'{e}}fan J.
                 van der Walt and Ralf Gommers and Pauli Virtanen and David
                 Cournapeau and Eric Wieser and Julian Taylor and Sebastian
                 Berg and Nathaniel J. Smith and Robert Kern and Matti Picus
                 and Stephan Hoyer and Marten H. van Kerkwijk and Matthew
                 Brett and Allan Haldane and Jaime Fern{\'{a}}ndez del
                 R{\'{i}}o and Mark Wiebe and Pearu Peterson and Pierre
                 G{\'{e}}rard-Marchant and Kevin Sheppard and Tyler Reddy and
                 Warren Weckesser and Hameer Abbasi and Christoph Gohlke and
                 Travis E. Oliphant},
 year          = {2020},
 month         = sep,
 journal       = {Nature},
 volume        = {585},
 number        = {7825},
 pages         = {357--362},
 doi           = {10.1038/s41586-020-2649-2},
 publisher     = {Springer Science and Business Media {LLC}},
 url           = {https://doi.org/10.1038/s41586-020-2649-2}
}

@ARTICLE{Garcia13,
       author = {{Garc{\'\i}a}, J. and {Dauser}, T. and {Reynolds}, C.~S. and {Kallman}, T.~R. and {McClintock}, J.~E. and {Wilms}, J. and {Eikmann}, W.},
        title = "{X-Ray Reflected Spectra from Accretion Disk Models. III. A Complete Grid of Ionized Reflection Calculations}",
      journal = {\apj},
         year = 2013,
        month = may,
       volume = {768},
       number = {2},
          eid = {146},
        pages = {146},
          doi = {10.1088/0004-637X/768/2/146},
archivePrefix = {arXiv},
       eprint = {1303.2112},
 primaryClass = {astro-ph.HE},
       adsurl = {https://ui.adsabs.harvard.edu/abs/2013ApJ...768..146G}
}

@ARTICLE{Dauser22,
       author = {{Dauser}, T. and {Garc{\'\i}a}, J.~A. and {Joyce}, A. and {Licklederer}, S. and {Connors}, R.~M.~T. and {Ingram}, A. and {Reynolds}, C.~S. and {Wilms}, J.},
        title = "{The effect of returning radiation on relativistic reflection}",
      journal = {\mnras},
         year = 2022,
        month = aug,
       volume = {514},
       number = {3},
        pages = {3965-3983},
          doi = {10.1093/mnras/stac1593},
archivePrefix = {arXiv},
       eprint = {2206.07973},
 primaryClass = {astro-ph.HE},
       adsurl = {https://ui.adsabs.harvard.edu/abs/2022MNRAS.514.3965D}
}

@ARTICLE{emcee,
   author = {{Foreman-Mackey}, D. and {Hogg}, D.~W. and {Lang}, D. and {Goodman}, J.
	},
    title = "{emcee: The MCMC Hammer}",
  journal = {\pasp},
archivePrefix = "arXiv",
   eprint = {1202.3665},
 primaryClass = "astro-ph.IM",
     year = 2013,
    month = mar,
   volume = 125,
    pages = {306},
      doi = {10.1086/670067},
   adsurl = {https://ui.adsabs.harvard.edu/abs/2013pasp..125..306F}
}

@INPROCEEDINGS{isis,
   author = {{Houck}, J.~C. and {Denicola}, L.~A.},
    title = "{ISIS: An Interactive Spectral Interpretation System for High Resolution X-Ray Spectroscopy}",
booktitle = {Astronomical Data Analysis Software and Systems IX},
     year = 2000,
   series = {Astronomical Society of the Pacific Conference Series},
   volume = 216,
   editor = {{Manset}, N. and {Veillet}, C. and {Crabtree}, D.},
    pages = {591},
   adsurl = {https://ui.adsabs.harvard.edu/abs/2000ASPC..216..591H}
}

@ARTICLE{nthcomp,
       author = {{{\.Z}ycki}, Piotr T. and {Done}, Chris and {Smith}, David A.},
       title = "{The 1989 May outburst of the soft X-ray transient GS 2023+338 (V404 Cyg)}",
      journal = {\mnras},
         year = "1999",
        month = "Nov",
       volume = {309},
       number = {3},
        pages = {561-575},
          doi = {10.1046/j.1365-8711.1999.02885.x},
archivePrefix = {arXiv},
       eprint = {astro-ph/9904304},
 primaryClass = {astro-ph},
       adsurl = {https://ui.adsabs.harvard.edu/abs/1999mnras.309..561Z}
}

@ARTICLE{diskbb,
       author = {{Makishima}, K. and {Maejima}, Y. and {Mitsuda}, K. and {Bradt}, H.~V. and
         {Remillard}, R.~A. and {Tuohy}, I.~R. and {Hoshi}, R. and
         {Nakagawa}, M.},
        title = "{Simultaneous X-Ray and Optical Observations of GX 339-4 in an X-Ray High State}",
      journal = {\apj},
         year = "1986",
        month = "Sep",
       volume = {308},
        pages = {635},
          doi = {10.1086/164534},
       adsurl = {https://ui.adsabs.harvard.edu/abs/1986apj...308..635M}
}

@ARTICLE{Buisson19,
       author = {{Buisson}, D.~J.~K. and {Fabian}, A.~C. and {Barret}, D. and {F{\"u}rst}, F. and {Gandhi}, P. and {Garc{\'\i}a}, J.~A. and {Kara}, E. and {Madsen}, K.~K. and {Miller}, J.~M. and {Parker}, M.~L. and {Shaw}, A.~W. and {Tomsick}, J.~A. and {Walton}, D.~J.},
        title = "{MAXI J1820+070 with NuSTAR I. An increase in variability frequency but a stable reflection spectrum: coronal properties and implications for the inner disc in black hole binaries}",
      journal = {\mnras},
         year = 2019,
        month = nov,
       volume = {490},
       number = {1},
        pages = {1350-1362},
          doi = {10.1093/mnras/stz2681},
archivePrefix = {arXiv},
       eprint = {1909.04688},
 primaryClass = {astro-ph.HE},
       adsurl = {https://ui.adsabs.harvard.edu/abs/2019MNRAS.490.1350B}
}

@ARTICLE{Ixpe1,
       author = {{Krawczynski}, Henric and {Muleri}, Fabio and {Dov{\v{c}}iak}, Michal and {Veledina}, Alexandra and {Rodriguez Cavero}, Nicole and {Svoboda}, Jiri and {Ingram}, Adam and {Matt}, Giorgio and {Garcia}, Javier A. and {Loktev}, Vladislav and {Negro}, Michela and {Poutanen}, Juri and {Kitaguchi}, Takao and {Podgorn{\'y}}, Jakub and {Rankin}, John and {Zhang}, Wenda and {Berdyugin}, Andrei and {Berdyugina}, Svetlana V. and {Bianchi}, Stefano and {Blinov}, Dmitry and {Capitanio}, Fiamma and {Di Lalla}, Niccol{\`o} and {Draghis}, Paul and {Fabiani}, Sergio and {Kagitani}, Masato and {Kravtsov}, Vadim and {Kiehlmann}, Sebastian and {Latronico}, Luca and {Lutovinov}, Alexander A. and {Mandarakas}, Nikos and {Marin}, Fr{\'e}d{\'e}ric and {Marinucci}, Andrea and {Miller}, Jon and {Mizuno}, Tsunefumi and {Molkov}, Sergey V. and {Omodei}, Nicola and {Petrucci}, Pierre-Olivier and {Ratheesh}, Ajay and {Sakanoi}, Takeshi and {Semena}, Andrei N. and {Skalidis}, Raphael and {Soffitta}, Paolo and {Tennant}, Allyn F. and {Thalhammer}, Phillipp and {Tombesi}, Francesco and {Weisskopf}, Martin C. and {Wilms}, Joern and {Zhang}, Sixuan and {Agudo}, Iv{\'a}n and {Antonelli}, Lucio A. and {Bachetti}, Matteo and {Baldini}, Luca and {Baumgartner}, Wayne H. and {Bellazzini}, Ronaldo and {Bongiorno}, Stephen D. and {Bonino}, Raffaella and {Brez}, Alessandro and {Bucciantini}, Niccol{\`o} and {Castellano}, Simone and {Cavazzuti}, Elisabetta and {Ciprini}, Stefano and {Costa}, Enrico and {De Rosa}, Alessandra and {Del Monte}, Ettore and {Di Gesu}, Laura and {Di Marco}, Alessandro and {Donnarumma}, Immacolata and {Doroshenko}, Victor and {Ehlert}, Steven R. and {Enoto}, Teruaki and {Evangelista}, Yuri and {Ferrazzoli}, Riccardo and {Gunji}, Shuichi and {Hayashida}, Kiyoshi and {Heyl}, Jeremy and {Iwakiri}, Wataru and {Jorstad}, Svetlana G. and {Karas}, Vladimir and {Kolodziejczak}, Jeffery J. and {La Monaca}, Fabio and {Liodakis}, Ioannis and {Maldera}, Simone and {Manfreda}, Alberto and {Marscher}, Alan P. and {Marshall}, Herman L. and {Massaro}, Francesco and {Mitsuishi}, Ikuyuki and {Ng}, C. -Y. and {O'Dell}, Stephen L. and {Oppedisano}, Chiara and {Papitto}, Alessandro and {Pavlov}, George G. and {Peirson}, Abel L. and {Perri}, Matteo and {Pesce-Rollins}, Melissa and {Pilia}, Maura and {Possenti}, Andrea and {Puccetti}, Simonetta and {Ramsey}, Brian D. and {Romani}, Roger W. and {Sgr{\`o}}, Carmelo and {Slane}, Patrick and {Spandre}, Gloria and {Tamagawa}, Toru and {Tavecchio}, Fabrizio and {Taverna}, Roberto and {Tawara}, Yuzuru and {Thomas}, Nicolas E. and {Trois}, Alessio and {Tsygankov}, Sergey and {Turolla}, Roberto and {Vink}, Jacco and {Wu}, Kinwah and {Xie}, Fei and {Zane}, Silvia},
        title = "{Polarized X-rays Constrain The Disk-Jet Geometry in a Black Hole X-ray Binary}",
      journal = {arXiv e-prints},
         year = 2022,
        month = jun,
          eid = {arXiv:2206.09972},
        pages = {arXiv:2206.09972},
archivePrefix = {arXiv},
       eprint = {2206.09972},
 primaryClass = {astro-ph.HE},
       adsurl = {https://ui.adsabs.harvard.edu/abs/2022arXiv220609972K}
}

@ARTICLE{Atri20,
       author = {{Atri}, P. and {Miller-Jones}, J.~C.~A. and {Bahramian}, A. and {Plotkin}, R.~M. and {Deller}, A.~T. and {Jonker}, P.~G. and {Maccarone}, T.~J. and {Sivakoff}, G.~R. and {Soria}, R. and {Altamirano}, D. and {Belloni}, T. and {Fender}, R. and {Koerding}, E. and {Maitra}, D. and {Markoff}, S. and {Migliari}, S. and {Russell}, D. and {Russell}, T. and {Sarazin}, C.~L. and {Tetarenko}, A.~J. and {Tudose}, V.},
        title = "{A radio parallax to the black hole X-ray binary MAXI J1820+070}",
      journal = {\mnras},
         year = 2020,
        month = mar,
       volume = {493},
       number = {1},
        pages = {L81-L86},
          doi = {10.1093/mnrasl/slaa010},
archivePrefix = {arXiv},
       eprint = {1912.04525},
 primaryClass = {astro-ph.HE},
       adsurl = {https://ui.adsabs.harvard.edu/abs/2020MNRAS.493L..81A}
}

@ARTICLE{Multinest1,
       author = {{Feroz}, F. and {Hobson}, M.~P.},
        title = "{Multimodal nested sampling: an efficient and robust alternative to Markov Chain Monte Carlo methods for astronomical data analyses}",
      journal = {\mnras},
         year = 2008,
        month = feb,
       volume = {384},
       number = {2},
        pages = {449-463},
          doi = {10.1111/j.1365-2966.2007.12353.x},
archivePrefix = {arXiv},
       eprint = {0704.3704},
 primaryClass = {astro-ph},
       adsurl = {https://ui.adsabs.harvard.edu/abs/2008MNRAS.384..449F}
}

@ARTICLE{Ultranest,
       author = {{Buchner}, Johannes},
        title = "{UltraNest - a robust, general purpose Bayesian inference engine}",
      journal = {The Journal of Open Source Software},
         year = 2021,
        month = apr,
       volume = {6},
       number = {60},
          eid = {3001},
        pages = {3001},
          doi = {10.21105/joss.03001},
archivePrefix = {arXiv},
       eprint = {2101.09604},
 primaryClass = {stat.CO},
       adsurl = {https://ui.adsabs.harvard.edu/abs/2021JOSS....6.3001B}
}

@ARTICLE{astropy13,
       author = {{Astropy Collaboration} and {Robitaille}, Thomas P. and {Tollerud}, Erik J. and {Greenfield}, Perry and {Droettboom}, Michael and {Bray}, Erik and {Aldcroft}, Tom and {Davis}, Matt and {Ginsburg}, Adam and {Price-Whelan}, Adrian M. and {Kerzendorf}, Wolfgang E. and {Conley}, Alexander and {Crighton}, Neil and {Barbary}, Kyle and {Muna}, Demitri and {Ferguson}, Henry and {Grollier}, Fr{\'e}d{\'e}ric and {Parikh}, Madhura M. and {Nair}, Prasanth H. and {Unther}, Hans M. and {Deil}, Christoph and {Woillez}, Julien and {Conseil}, Simon and {Kramer}, Roban and {Turner}, James E.~H. and {Singer}, Leo and {Fox}, Ryan and {Weaver}, Benjamin A. and {Zabalza}, Victor and {Edwards}, Zachary I. and {Azalee Bostroem}, K. and {Burke}, D.~J. and {Casey}, Andrew R. and {Crawford}, Steven M. and {Dencheva}, Nadia and {Ely}, Justin and {Jenness}, Tim and {Labrie}, Kathleen and {Lim}, Pey Lian and {Pierfederici}, Francesco and {Pontzen}, Andrew and {Ptak}, Andy and {Refsdal}, Brian and {Servillat}, Mathieu and {Streicher}, Ole},
        title = "{Astropy: A community Python package for astronomy}",
      journal = {\aap},
         year = 2013,
        month = oct,
       volume = {558},
          eid = {A33},
        pages = {A33},
          doi = {10.1051/0004-6361/201322068},
archivePrefix = {arXiv},
       eprint = {1307.6212},
 primaryClass = {astro-ph.IM},
       adsurl = {https://ui.adsabs.harvard.edu/abs/2013A&A...558A..33A}
}

@ARTICLE{astropy18,
       author = {{Astropy Collaboration} and {Price-Whelan}, A.~M. and {Sip{\H{o}}cz}, B.~M. and {G{\"u}nther}, H.~M. and {Lim}, P.~L. and {Crawford}, S.~M. and {Conseil}, S. and {Shupe}, D.~L. and {Craig}, M.~W. and {Dencheva}, N. and {Ginsburg}, A. and {VanderPlas}, J.~T. and {Bradley}, L.~D. and {P{\'e}rez-Su{\'a}rez}, D. and {de Val-Borro}, M. and {Aldcroft}, T.~L. and {Cruz}, K.~L. and {Robitaille}, T.~P. and {Tollerud}, E.~J. and {Ardelean}, C. and {Babej}, T. and {Bach}, Y.~P. and {Bachetti}, M. and {Bakanov}, A.~V. and {Bamford}, S.~P. and {Barentsen}, G. and {Barmby}, P. and {Baumbach}, A. and {Berry}, K.~L. and {Biscani}, F. and {Boquien}, M. and {Bostroem}, K.~A. and {Bouma}, L.~G. and {Brammer}, G.~B. and {Bray}, E.~M. and {Breytenbach}, H. and {Buddelmeijer}, H. and {Burke}, D.~J. and {Calderone}, G. and {Cano Rodr{\'\i}guez}, J.~L. and {Cara}, M. and {Cardoso}, J.~V.~M. and {Cheedella}, S. and {Copin}, Y. and {Corrales}, L. and {Crichton}, D. and {D'Avella}, D. and {Deil}, C. and {Depagne}, {\'E}. and {Dietrich}, J.~P. and {Donath}, A. and {Droettboom}, M. and {Earl}, N. and {Erben}, T. and {Fabbro}, S. and {Ferreira}, L.~A. and {Finethy}, T. and {Fox}, R.~T. and {Garrison}, L.~H. and {Gibbons}, S.~L.~J. and {Goldstein}, D.~A. and {Gommers}, R. and {Greco}, J.~P. and {Greenfield}, P. and {Groener}, A.~M. and {Grollier}, F. and {Hagen}, A. and {Hirst}, P. and {Homeier}, D. and {Horton}, A.~J. and {Hosseinzadeh}, G. and {Hu}, L. and {Hunkeler}, J.~S. and {Ivezi{\'c}}, {\v{Z}}. and {Jain}, A. and {Jenness}, T. and {Kanarek}, G. and {Kendrew}, S. and {Kern}, N.~S. and {Kerzendorf}, W.~E. and {Khvalko}, A. and {King}, J. and {Kirkby}, D. and {Kulkarni}, A.~M. and {Kumar}, A. and {Lee}, A. and {Lenz}, D. and {Littlefair}, S.~P. and {Ma}, Z. and {Macleod}, D.~M. and {Mastropietro}, M. and {McCully}, C. and {Montagnac}, S. and {Morris}, B.~M. and {Mueller}, M. and {Mumford}, S.~J. and {Muna}, D. and {Murphy}, N.~A. and {Nelson}, S. and {Nguyen}, G.~H. and {Ninan}, J.~P. and {N{\"o}the}, M. and {Ogaz}, S. and {Oh}, S. and {Parejko}, J.~K. and {Parley}, N. and {Pascual}, S. and {Patil}, R. and {Patil}, A.~A. and {Plunkett}, A.~L. and {Prochaska}, J.~X. and {Rastogi}, T. and {Reddy Janga}, V. and {Sabater}, J. and {Sakurikar}, P. and {Seifert}, M. and {Sherbert}, L.~E. and {Sherwood-Taylor}, H. and {Shih}, A.~Y. and {Sick}, J. and {Silbiger}, M.~T. and {Singanamalla}, S. and {Singer}, L.~P. and {Sladen}, P.~H. and {Sooley}, K.~A. and {Sornarajah}, S. and {Streicher}, O. and {Teuben}, P. and {Thomas}, S.~W. and {Tremblay}, G.~R. and {Turner}, J.~E.~H. and {Terr{\'o}n}, V. and {van Kerkwijk}, M.~H. and {de la Vega}, A. and {Watkins}, L.~L. and {Weaver}, B.~A. and {Whitmore}, J.~B. and {Woillez}, J. and {Zabalza}, V. and {Astropy Contributors}},
        title = "{The Astropy Project: Building an Open-science Project and Status of the v2.0 Core Package}",
      journal = {\aj},
         year = 2018,
        month = sep,
       volume = {156},
       number = {3},
          eid = {123},
        pages = {123},
          doi = {10.3847/1538-3881/aabc4f},
archivePrefix = {arXiv},
       eprint = {1801.02634},
 primaryClass = {astro-ph.IM},
       adsurl = {https://ui.adsabs.harvard.edu/abs/2018AJ....156..123A}
}

@ARTICLE{astropy22,
       author = {{Astropy Collaboration} and {Price-Whelan}, Adrian M. and {Lim}, Pey Lian and {Earl}, Nicholas and {Starkman}, Nathaniel and {Bradley}, Larry and {Shupe}, David L. and {Patil}, Aarya A. and {Corrales}, Lia and {Brasseur}, C.~E. and {N{"o}the}, Maximilian and {Donath}, Axel and {Tollerud}, Erik and {Morris}, Brett M. and {Ginsburg}, Adam and {Vaher}, Eero and {Weaver}, Benjamin A. and {Tocknell}, James and {Jamieson}, William and {van Kerkwijk}, Marten H. and {Robitaille}, Thomas P. and {Merry}, Bruce and {Bachetti}, Matteo and {G{"u}nther}, H. Moritz and {Aldcroft}, Thomas L. and {Alvarado-Montes}, Jaime A. and {Archibald}, Anne M. and {B{'o}di}, Attila and {Bapat}, Shreyas and {Barentsen}, Geert and {Baz{'a}n}, Juanjo and {Biswas}, Manish and {Boquien}, M{'e}d{'e}ric and {Burke}, D.~J. and {Cara}, Daria and {Cara}, Mihai and {Conroy}, Kyle E. and {Conseil}, Simon and {Craig}, Matthew W. and {Cross}, Robert M. and {Cruz}, Kelle L. and {D'Eugenio}, Francesco and {Dencheva}, Nadia and {Devillepoix}, Hadrien A.~R. and {Dietrich}, J{"o}rg P. and {Eigenbrot}, Arthur Davis and {Erben}, Thomas and {Ferreira}, Leonardo and {Foreman-Mackey}, Daniel and {Fox}, Ryan and {Freij}, Nabil and {Garg}, Suyog and {Geda}, Robel and {Glattly}, Lauren and {Gondhalekar}, Yash and {Gordon}, Karl D. and {Grant}, David and {Greenfield}, Perry and {Groener}, Austen M. and {Guest}, Steve and {Gurovich}, Sebastian and {Handberg}, Rasmus and {Hart}, Akeem and {Hatfield-Dodds}, Zac and {Homeier}, Derek and {Hosseinzadeh}, Griffin and {Jenness}, Tim and {Jones}, Craig K. and {Joseph}, Prajwel and {Kalmbach}, J. Bryce and {Karamehmetoglu}, Emir and {Ka{l}uszy{'n}ski}, Miko{l}aj and {Kelley}, Michael S.~P. and {Kern}, Nicholas and {Kerzendorf}, Wolfgang E. and {Koch}, Eric W. and {Kulumani}, Shankar and {Lee}, Antony and {Ly}, Chun and {Ma}, Zhiyuan and {MacBride}, Conor and {Maljaars}, Jakob M. and {Muna}, Demitri and {Murphy}, N.~A. and {Norman}, Henrik and {O'Steen}, Richard and {Oman}, Kyle A. and {Pacifici}, Camilla and {Pascual}, Sergio and {Pascual-Granado}, J. and {Patil}, Rohit R. and {Perren}, Gabriel I. and {Pickering}, Timothy E. and {Rastogi}, Tanuj and {Roulston}, Benjamin R. and {Ryan}, Daniel F. and {Rykoff}, Eli S. and {Sabater}, Jose and {Sakurikar}, Parikshit and {Salgado}, Jes{'u}s and {Sanghi}, Aniket and {Saunders}, Nicholas and {Savchenko}, Volodymyr and {Schwardt}, Ludwig and {Seifert-Eckert}, Michael and {Shih}, Albert Y. and {Jain}, Anany Shrey and {Shukla}, Gyanendra and {Sick}, Jonathan and {Simpson}, Chris and {Singanamalla}, Sudheesh and {Singer}, Leo P. and {Singhal}, Jaladh and {Sinha}, Manodeep and {Sip{H{o}}cz}, Brigitta M. and {Spitler}, Lee R. and {Stansby}, David and {Streicher}, Ole and {{{S}}umak}, Jani and {Swinbank}, John D. and {Taranu}, Dan S. and {Tewary}, Nikita and {Tremblay}, Grant R. and {Val-Borro}, Miguel de and {Van Kooten}, Samuel J. and {Vasovi{'c}}, Zlatan and {Verma}, Shresth and {de Miranda Cardoso}, Jos{'e} Vin{'i}cius and {Williams}, Peter K.~G. and {Wilson}, Tom J. and {Winkel}, Benjamin and {Wood-Vasey}, W.~M. and {Xue}, Rui and {Yoachim}, Peter and {Zhang}, Chen and {Zonca}, Andrea and {Astropy Project Contributors}},
        title = "{The Astropy Project: Sustaining and Growing a Community-oriented Open-source Project and the Latest Major Release (v5.0) of the Core Package}",
      journal = {\apj},
         year = 2022,
        month = aug,
       volume = {935},
       number = {2},
          eid = {167},
        pages = {167},
          doi = {10.3847/1538-4357/ac7c74},
archivePrefix = {arXiv},
       eprint = {2206.14220},
 primaryClass = {astro-ph.IM},
       adsurl = {https://ui.adsabs.harvard.edu/abs/2022ApJ...935..167A}
}

@ARTICLE{Buchner23b,
       author = {{Buchner}, Johannes},
        title = "{Nested Sampling Methods}",
      journal = {Statistics Surveys},
         year = 2023,
        month = jan,
       volume = {17},
        pages = {169-215},
          doi = {10.1214/23-SS144},
archivePrefix = {arXiv},
       eprint = {2101.09675},
 primaryClass = {stat.CO},
       adsurl = {https://ui.adsabs.harvard.edu/abs/2023StSur..17..169B}
}

@ARTICLE{Poutanen96,
       author = {{Poutanen}, Juri and {Svensson}, Roland},
        title = "{The Two-Phase Pair Corona Model for Active Galactic Nuclei and X-Ray Binaries: How to Obtain Exact Solutions}",
      journal = {\apj},
         year = 1996,
        month = oct,
       volume = {470},
        pages = {249},
          doi = {10.1086/177865},
archivePrefix = {arXiv},
       eprint = {astro-ph/9605073},
 primaryClass = {astro-ph},
       adsurl = {https://ui.adsabs.harvard.edu/abs/1996ApJ...470..249P}
}

@ARTICLE{Dauser13,
       author = {{Dauser}, T. and {Garcia}, J. and {Wilms}, J. and {B{\"o}ck}, M. and
         {Brenneman}, L.~W. and {Falanga}, M. and {Fukumura}, K. and
         {Reynolds}, C.~S.},
        title = "{Irradiation of an accretion disc by a jet: general properties and implications for spin measurements of black holes}",
      journal = {\mnras},
         year = 2013,
        month = apr,
       volume = {430},
       number = {3},
        pages = {1694-1708},
          doi = {10.1093/mnras/sts710},
archivePrefix = {arXiv},
       eprint = {1301.4922},
 primaryClass = {astro-ph.HE},
       adsurl = {https://ui.adsabs.harvard.edu/abs/2013MNRAS.430.1694D}
}

@ARTICLE{Garcia10,
       author = {{Garc{\'\i}a}, J. and {Kallman}, T.~R.},
        title = "{X-ray Reflected Spectra from Accretion Disk Models. I. Constant Density Atmospheres}",
      journal = {\apj},
         year = 2010,
        month = aug,
       volume = {718},
       number = {2},
        pages = {695-706},
          doi = {10.1088/0004-637X/718/2/695},
archivePrefix = {arXiv},
       eprint = {1006.0485},
 primaryClass = {astro-ph.HE},
       adsurl = {https://ui.adsabs.harvard.edu/abs/2010ApJ...718..695G}
}

@INPROCEEDINGS{Arnaud96,
       author = {{Arnaud}, K.~A.},
        title = "{XSPEC: The First Ten Years}",
    booktitle = {Astronomical Data Analysis Software and Systems V},
         year = 1996,
       editor = {{Jacoby}, George H. and {Barnes}, Jeannette},
       series = {Astronomical Society of the Pacific Conference Series},
       volume = {101},
        month = jan,
        pages = {17},
       adsurl = {https://ui.adsabs.harvard.edu/abs/1996ASPC..101...17A}
}

@ARTICLE{jaxspec,
       author = {{Dupourqu{\'e}}, S. and {Barret}, D. and {Diez}, C.~M. and {Guillot}, S. and {Quintin}, E.},
        title = "{jaxspec: A fast and robust Python library for X-ray spectral fitting}",
      journal = {\aap},
         year = 2024,
        month = oct,
       volume = {690},
          eid = {A317},
        pages = {A317},
          doi = {10.1051/0004-6361/202451736},
       adsurl = {https://ui.adsabs.harvard.edu/abs/2024A&A...690A.317D}
}

@ARTICLE{Garcia15,
       author = {{Garc{\'\i}a}, Javier A. and {Steiner}, James F. and
         {McClintock}, Jeffrey E. and {Remillard}, Ronald A. and
         {Grinberg}, Victoria and {Dauser}, Thomas},
        title = "{X-Ray Reflection Spectroscopy of the Black Hole GX 339--4: Exploring the Hard State with Unprecedented Sensitivity}",
      journal = {\apj},
         year = 2015,
        month = nov,
       volume = {813},
       number = {2},
          eid = {84},
        pages = {84},
          doi = {10.1088/0004-637X/813/2/84},
archivePrefix = {arXiv},
       eprint = {1505.03607},
 primaryClass = {astro-ph.HE},
       adsurl = {https://ui.adsabs.harvard.edu/abs/2015ApJ...813...84G}
}
\bibliographystyle{aa}

%%%%%%%%%%%%%%%%%%%%%%%%%%%%%%%%%%%%%%%%%%%%%%%%%%%%%%%%%%%%%%%
% Appendices must be placed after   \end{thebibliography}
% They will be placed automatically on a new page.
%%%%%%%%%%%%%%%%%%%%%%%%%%%%%%%%%%%%%%%%%%%%%%%%%%%%%%%%%%%%%%%

\begin{appendix} 

\section{One-dimensional interpolation error}
\label{sec:1d_error}
Given a one-parameter model $m(\theta)$ and its tabulated version $\tilde{m}(\theta)$, the interpolation error is
\begin{equation}
    e(\theta) = m(\theta)-\tilde{m}(\theta),
\end{equation}
which by definition is $0$ at the grid points $\theta_i$ and $\theta_{i+1}$. We want to compute the error for any other point $\theta$ between the grid points. Let $w(t) = -(t-\theta_{i})(\theta_{i+1}-t)$, where $t \in [\theta_i,\theta_{i+1}]$ is also contained between the two grid points. We now define the function $g_\theta(t)$:
\begin{equation}
    g_\theta(t) = e(t) - \frac{e(\theta)}{w(\theta)}w(t)
\end{equation}
This function vanishes (at least) at the points $\theta_i$, $\theta$, and $\theta_{i+1}$. As a result, its first derivative with respect to $t$, $d g_\theta(t)/dt$ vanishes at (at least) two points $\in [\theta_i,\theta_{i+1}]$, and $d^{2} g_\theta(t)/dt^{2}$ vanishes at (at least) one point. $d^{2}\tilde{m}(t)/dt^{2}=0$ because by definition $\tilde{m}$ is linear. Trivially, $d^{2} w(t)/dt^{2}=2$. The second derivative $d^{2} g_\theta(t)/dt^{2}$ then is:
\begin{equation}
    \frac{d^{2} g_\theta(t)}{dt^{2}} = \frac{d^{2} m(t)}{dt^{2}} - 2\frac{e(\theta)}{w(\theta)};
\end{equation}
We know by Rolle's theorem that there is (at least) one point $t=\xi \in [\theta_i,\theta_{i+1}]$ such that the second derivative vanishes, $d^{2} g_\theta(t)/dt^{2}=0$. Therefore, evaluating the equation at this particular point $\xi$, simplifying the derivative notation, and substituting the definition of $w(\theta)$, we have:
\begin{equation}
    0 = m^{\prime\prime}(\xi) + 2\frac{e(\theta)}{(\theta-\theta_{i})(\theta_{i+1}-\theta)},
\end{equation}
from which we can trivially solve for $e(\theta)$:
\begin{equation}
    e(\theta) = -\frac{1}{2}m^{\prime\prime}(\xi) (\theta-\theta_{i})(\theta_{i+1}-\theta).
\end{equation}

\section{Likelihood from interpolating multiple parameters}
\label{sec:nd_derivation}
Given a model $m(\bm{\theta})$, we can write its second order Taylor expansion a distance $\bm{\theta}$ away from the lower bound of the grid cell $\bm{c}$ as:
\begin{equation}
    m(\bm{\theta}) = m(\bm{c}) - \sum_j m^{\prime}_j(\bm{c}) u_j + \frac{1}{2}\sum_{j\neq l} H_{j,l}u_ju_l + \frac{1}{2}\sum_{j} H_{j,j}u_j^{2},
    \label{eq:nd_expansion}
\end{equation}
where $H_{j,l}$ is the Hessian matrix of the model and $u_j(\theta_j) = c_j-\theta_j$. We can then write the same formula for the table model:
\begin{equation}
    \tilde{m}(\bm{\theta}) = m(\bm{c}) - \sum_j m^{\prime}_j(\bm{c}) u_j + \frac{1}{2}\sum_{j\neq l} H_{j,l}u_ju_l + \frac{1}{2}\sum_{j} H_{j,j}\tilde{u}_j^{2},   
\end{equation}
crucially, the constant term $m(\bm{c})$ as well as the first order terms in $u_j$ are identical to eq.\ref{eq:nd_expansion} because by definition linear interpolation reproduces them exactly. $\tilde{u}_j^{2}$ is also estimated through linear interpolation, rather than being computed exactly from the model. Therefore, to second order, the interpolation error depends on how accurately these second order terms are estimated:
\begin{equation}
    e_k (\bm{\theta}) = \frac{1}{2}\sum_j H_{j,j} (u_j^{2}-\tilde{u}_j^{2}).
\end{equation}
We now need to write the interpolant function $u_j^{2}(\theta_j)-\tilde{u}_j^{2}(\theta_j)$ explicitly. Because it depends exclusively on the $j$-th parameter, it is identical to the one-dimensional form:
\begin{equation}
    u_j^{2}(\theta_j)-\tilde{u}_j^{2}(\theta_j) = -(\theta_j-\theta_{j,i})(\theta_{j,i+1}-\theta_{j}),
\end{equation}
and substituting we immediately re-derive eq.\ref{eq:nd_error}:
\begin{equation}
 e (\bm{\theta}) = -\frac{1}{2}\sum_j H_{j,j} (\theta_j-\theta_{j,i})(\theta_{j,i+1}-\theta_{j})   
\end{equation}

Eqs.\ref{eq:nd_likelihood}-\ref{eq:nd_deriv} then follow immediately, by replacing the single interpolation error $e(\theta)$ of the one-dimensional case with the sum of the n-dimensional errors $e(\bm{\theta})$.

\section{Logarithmic Interpolation} 
\label{sec:log_interp}

\begin{figure}
    \centering
    \includegraphics[width=\linewidth]{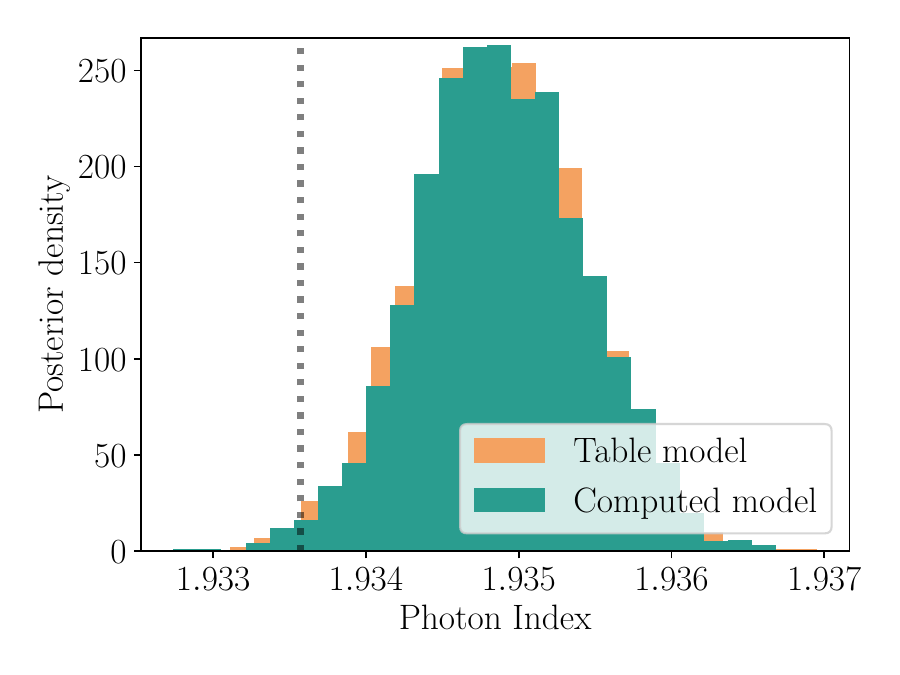}
    \caption{Left panel of Fig. \ref{fig:pl_posterior} when using logarithmic interpolation.}
    \label{fig:log_interp_chi}
\end{figure}

The natural question to ask when considering the pitfalls of the use of interpolation is what possible alternatives exist. For the power law case, the easiest solution is logarithmic interpolation as a power law is linear in logarithmic space. In Fig \ref{fig:log_interp_chi}, we show that, when using logarithmic interpolation, the posteriors between the computed model are largely in agreement. One may naively think that the solution to the issue is simply to adopt logarithmic interpolation. The issue arises that many parameters within a table model will not produce logarithmic changes in the flux at any given energy; this will introduce even further unrealistic complexity into the likelihood.

\section{Model posteriors}
\begin{figure*}
    \centering
    \includegraphics[width=\textwidth]{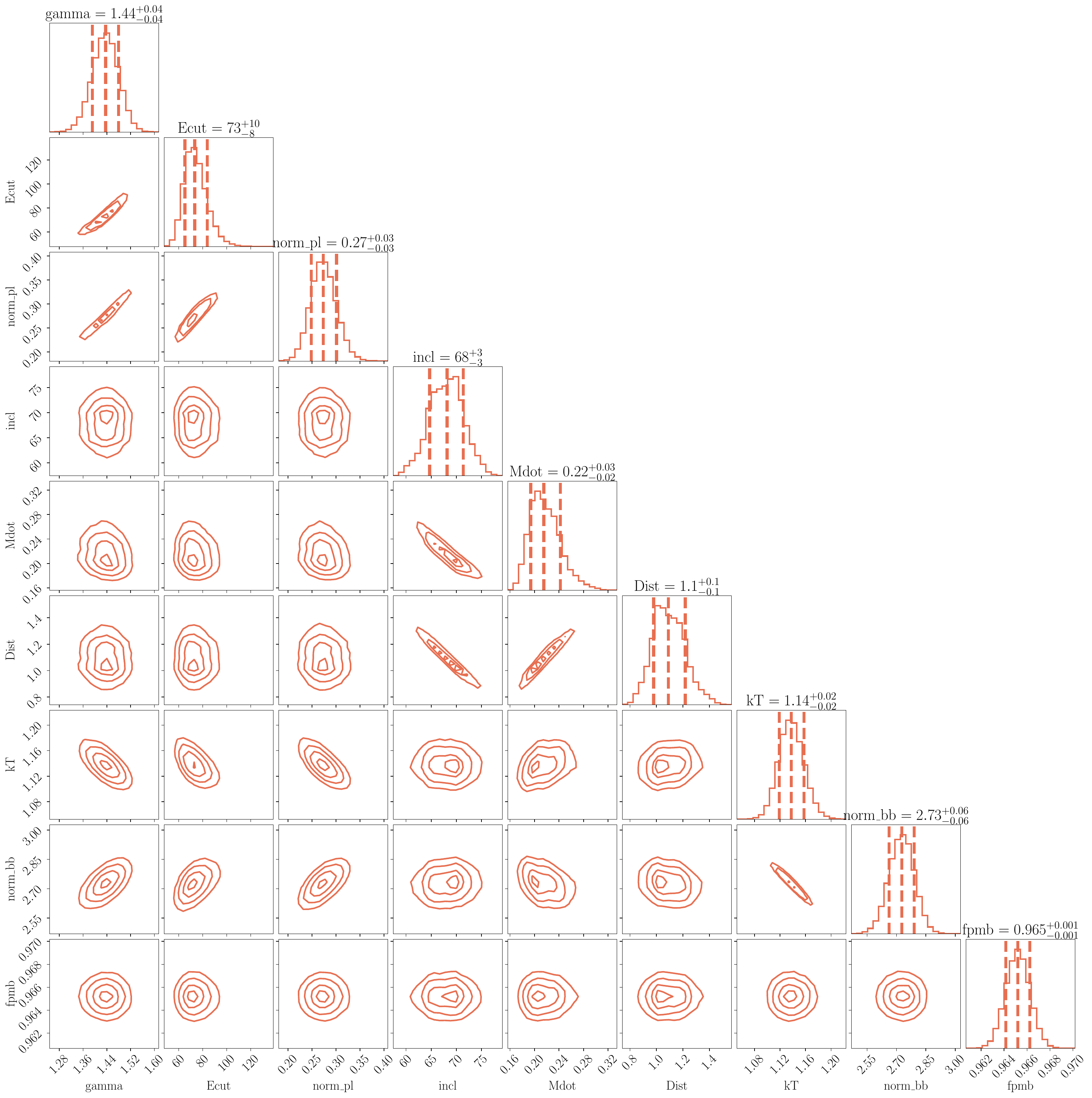}
    \caption{One and two dimensional posteriors for the SIMS fit with spin fixed to $a=0.75$. The dashed lines indicate the 68th-percentile of each distribution.}
    \label{fig:posterior_fixspin}
\end{figure*}

\begin{figure*}
    \centering
    \includegraphics[width=\textwidth]{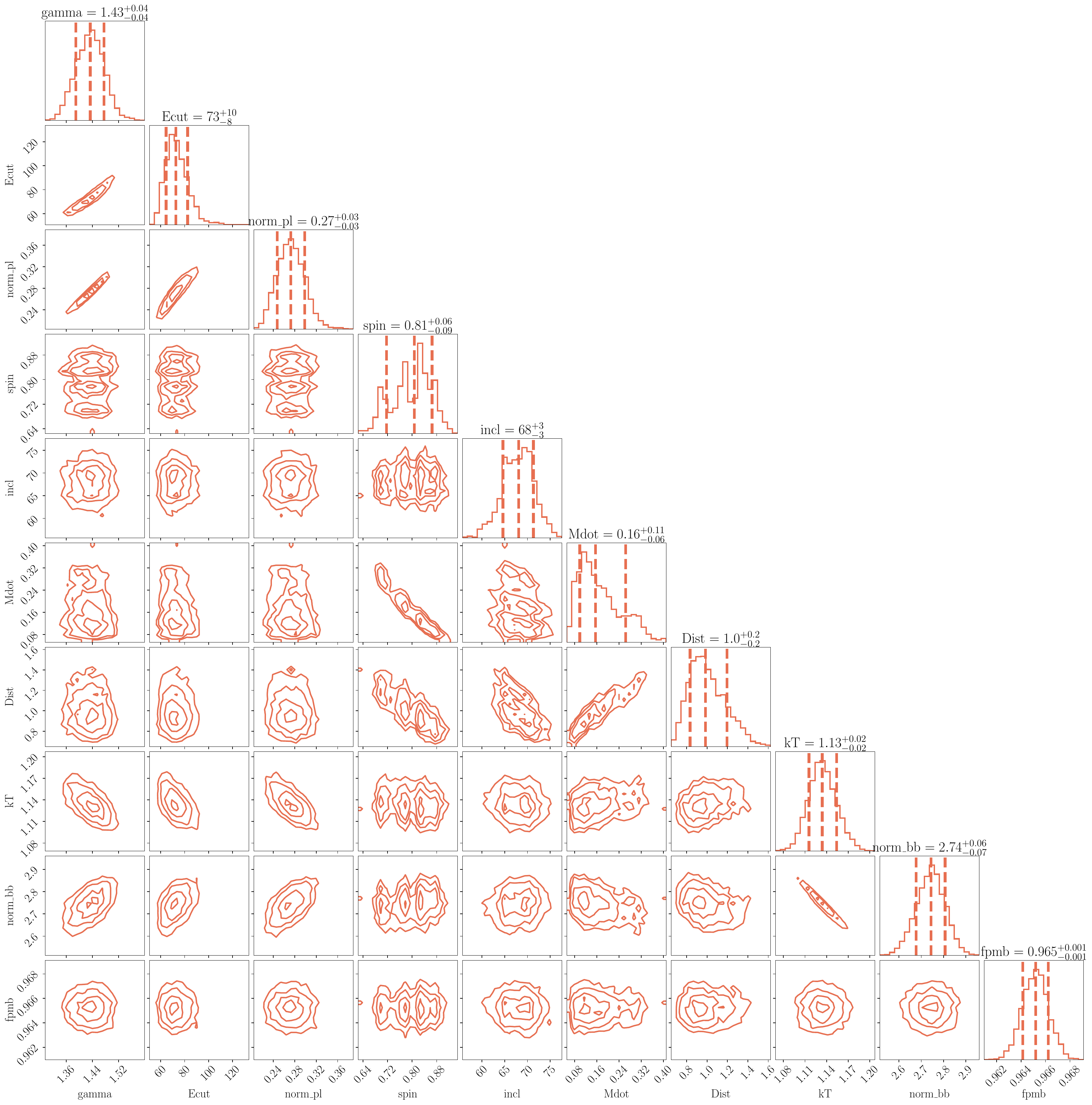}
    \caption{One and two dimensional posteriors for the SIMS fit with free spin. The dashed lines indicate the 68th-percentile of each distribution.}
    \label{fig:posterior_freespin}
\end{figure*}

\begin{figure*}
    \centering
    \includegraphics[width=\textwidth]{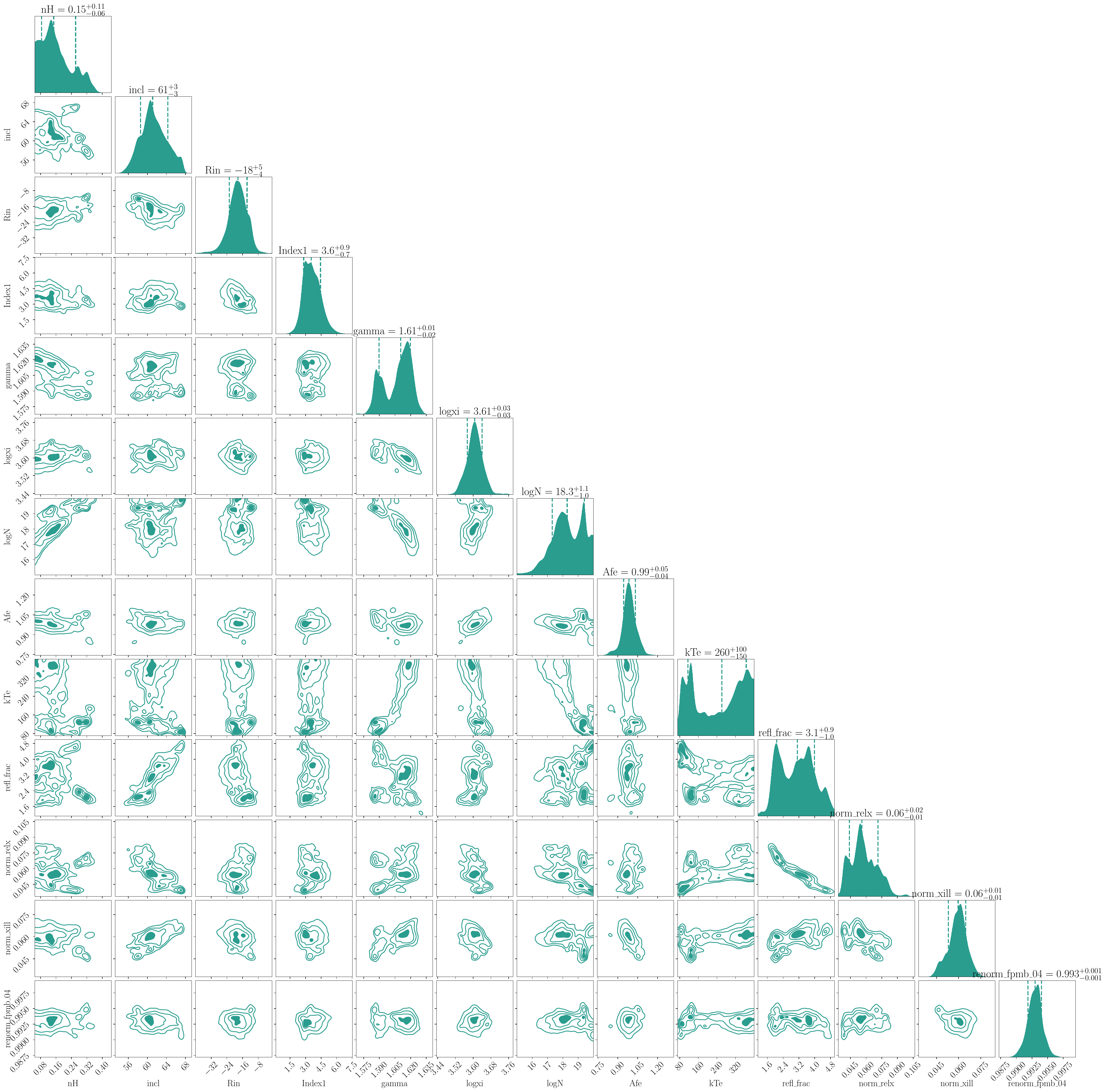}
    \caption{One and two dimensional posteriors for the hard state fit. The dashed lines indicate the 68th-percentile of each distribution.}
    \label{fig:posterior_reflection}
\end{figure*}

In this section we report the full model posteriors from the fits discussed in Sec.\ref{sec:softstate} and \ref{sec:hardstate}. Fig.\ref{fig:posterior_fixspin} shows the posterior distributions from the SIMS fit with fixed spin. Fig.\ref{fig:posterior_freespin} shows the posterior distributions from the SIMS fit with free spin. Fig.\ref{fig:posterior_reflection} shows the posterior distributions from the hard state fit. 

\end{appendix}

\end{document}